\documentclass[12pt]{article}
\usepackage{amsmath}

\usepackage{times}
\usepackage{bm}
\usepackage{natbib}
\usepackage{graphicx}
\usepackage{caption,subcaption}
\usepackage{float}
\usepackage{epstopdf}
\usepackage{color,comment}
\usepackage[dvipsnames]{xcolor}
\usepackage{pdflscape}
\usepackage{setspace,amsfonts, amssymb,array}
\usepackage{geometry}
\usepackage{authblk}
\newcolumntype{H}{>{\setbox0=\hbox\bgroup}c<{\egroup}@{}}
\newtheorem{theorem}{Theorem}
\newtheorem{lemma}{Lemma}

\usepackage[plain,noend]{algorithm2e}
\makeatletter
\renewcommand{\algocf@captiontext}[2]{#1\algocf@typo. \AlCapFnt{}#2} 
\def\@algocf@capt@plain{top}
\renewcommand{\algocf@makecaption}[2]{%
	\addtolength{\hsize}{\algomargin}%
	\sbox\@tempboxa{\algocf@captiontext{#1}{#2}}%
	\ifdim\wd\@tempboxa >\hsize
	\hskip .5\algomargin%
	\parbox[t]{\hsize}{\algocf@captiontext{#1}{#2}}
	\else%
	\global\@minipagefalse%
	\hbox to\hsize{\box\@tempboxa}
	\fi%
	\addtolength{\hsize}{-\algomargin}%
}
\makeatother

\title{Estimation in linear errors-in-variables models \\with unknown error distribution}

\author[1]{L. H. Nghiem \thanks{Email: lnghiem@smu.edu}}
\author[1]{M. C. Byrd \thanks{Email: mcbyrd@smu.edu}} 
\author[1,2]{C. J. Potgiter \thanks{Email: cjpotgieter@smu.edu}} 
\affil[1]{Department of Statistical Science, Southern Methodist University, Dallas, TX}
\affil[2]{Department of Statistics, University of Johannesburg, South Africa}

\begin{document}

\maketitle

\begin{abstract}
	Parameter estimation in linear errors-in-variables models typically requires that the measurement error distribution be known (or estimable from replicate data). A generalized method of moments approach can be used to estimate model parameters in the absence of knowledge of the error distributions, but requires the existence of a large number of model moments. In this paper, parameter estimation based on the phase function, a normalized version of the characteristic function, is considered. This approach requires the model covariates to have asymmetric distributions, while the error distributions are symmetric. Parameter estimation is then based on minimizing a distance function between the empirical phase functions of the noisy covariates and the outcome variable. No knowledge of the measurement error distribution is required to calculate this estimator. Both the asymptotic and finite sample properties of the estimator are considered. The connection between the phase function approach and method of moments is also discussed. The estimation of standard errors is also considered and a modified bootstrap algorithm is proposed for fast computation. The newly proposed estimator is competitive when compared to generalized method of moments, even while making fewer model assumptions on the measurement error. Finally, the proposed method is applied to a real dataset concerning the measurement of air pollution.
\end{abstract}

Keywords: Asymmetric distribution; Empirical characteristic function; Measurement error; Method of moments; Phase function.

\section{Introduction}
Errors-in-variables models arise when some covariates cannot be measured accurately. Sources of measurement error include the instruments used to measure the variables of interest and the inadequacy of measurements taken over the short term being used as proxies for long-term variables. In the classic measurement error framework, this results in observed covariates having larger variance than the true predictors. Let $X = (X^{(1)},\ldots,X^{(p)})^\top \in \mathbb{R}^{p}$ denote the true model covariates and let $Y \in \mathbb{R}$ denote the outcome of interest. For $\beta_1 \in \mathbb{R}^{p}$, the relationship between $X$ and $Y$ is assumed to be $Y = \beta_0 + X^\top \beta_1 + \varepsilon$ with intercept $\beta_0 \in \mathbb{R}$ and error $\varepsilon\in \mathbb{R}$. In an errors-in-variables model, $X$ is not directly observed. Rather, $W=(W^{(1)},\ldots,W^{(p)})^\top \in \mathbb{R}^{p}$ is observed with $W= X + U$ denoting the covariates contaminated by additive measurement error, and $U \in \mathbb{R}^{p}$ denoting the measurement error. This model represents the classic formulation of the errors-in-variables model and the estimation of $\beta=(\beta_0,\beta_1^\top)^\top$ is of interest.

Above, the model error $\varepsilon$ is assumed to be symmetric about $0$ with scale parameter $\sigma^2$ and the measurement error $U$ is assumed to be symmetric about $0\in \mathbb{R}^{p}$ with scale matrix $\Sigma_u$. Generally, $\sigma^2$ and $\Sigma_u$ represent, respectively, the variance of $\varepsilon$ and covariance matrix of $U$ when these quantities are well-defined. The covariates $X$, measurement error $U$ and model error $\varepsilon$ are furthermore assumed mutually independent. Given a sample $(W_1,Y_1),\ldots,(W_n,Y_n)$, it is well known that regression of the $Y_i$ on the $W_i$ using traditional methods such as ordinary least squares leads to an inconsistent and biased estimate of $\beta$, see \cite{carroll2006measurement}. Hence, adjusting for the presence of measurement error is important for accurately describing the relationship between the true covariates and the outcome of interest. 

There is a vast literature on correcting for the effect of measurement error dating back to \cite{wald1940fitting}. An overview of more modern approaches to estimation in the presence of measurement error can be found in \cite{fuller2009measurement}. Most of the methods proposed in the literature require strong assumptions on the distribution of the measurement error $U$. Specifically, it is generally assumed that the covariance matrix $\Sigma_u$ is known. For example, in the simple linear model setting where $Y= \beta_0 + \beta_1 X + \varepsilon$ and $ W = X + U$, a consistent estimator of the slope $\beta_1$ is obtained as $\hat{\beta_{1}}=\hat{\beta}^{(W)}_1\sigma^2_X/(\sigma^2_X + \sigma^2_U)$, where $\hat{\beta}^{(W)}_1$ is the estimated slope obtained by ordinary least squares regression of $Y$ on $W$. Calculation of this estimator requires that the variance of measurement error $\sigma^2_U$ be known or estimable.  In the multiple predictor setting, there is generally no closed-form solution for the corrected estimator. Instead, simulation-extrapolation , first proposed by \cite{stefanski1995simulation}, is frequently used. The simulation-extrapolation  procedure evaluates the effects of measurement error on the estimator by increasing the level of measurement error through a simulation step, and then extrapolating to the setting of no measurement error. simulation-extrapolation  also requires that $\Sigma_u$ be known or estimable. In many commonly assumed scenarios, such as when both the covariates $X$ and measurement error $U$ are assumed to be multivariate normal, $\Sigma_u$ can only be estimated in the presence of auxiliary data, see \cite{reiersol1950identifiability} for a discussion of the identifiability of errors-in-variables models. Another approach to correcting for measurement error is regression calibration, see \cite{carroll1990approximate}. Here, a regression of $X$ on $W$ is used to estimate $X$, say $\hat{X}$, and then the linear model parameters are estimated by regressing $Y$ on $\hat{X}$. The regression of $X$ on $W$ is assumed to be available through an either validation data or an instrumental variable $T$. 

When the distributions of $X$, $U$ and $\varepsilon$ are fully specified, likelihood methods can also be used to estimate parameters, see \cite{schafer1996likelihood} and \cite{higdon2001maximum}. Implementation of these likelihood methods generally require the use of numerical methods such as Gaussian quadrature or Monte Carlo integration. The EM algorithm of \cite{dempster1977maximum} can also be used. An approach that does not require the distribution of $X$ to be known is the conditional score method of \cite{stefanski1987conditional}. However, this method does require that both parametric models for the conditional distributions $Y\mid X$ and $W\mid X$ be specified. 

The method of moments approach to estimating linear errors-in-variables models dates back to the work of \cite{reiersol1941confluence}, who estimates the slope of the simple errors-in-variables model through third-order moments. \cite{gillard2014method} considers slope estimators based on third and fourth moments,  and finds these to have large variances. More recently, methods based on the matching of higher-order moments, or variants such as cumulants, have been explored with renewed interest. \cite{erickson2002two} express high-order residual moments as nonlinear functions of both coefficients and nuisance parameters, while \cite{erickson2014minimum} express the third and fourth residual cumulants as a linear function of the coefficients. The latter also establishes that the two methods are asymptotically equivalent. The method of moments approach is nonparametric in the sense that it does not require parametric distributions to be specified for any of the components. However, an implementation based on the first $M$ sample moments generally requires $2M$ finite population moments.

This paper proposes a method of estimation that is fully nonparametric, in that implementation does not require parametric specifications of any model components, nor does it require the existence of model moments. Furthermore, the method does not require that the measurement error variance be known, if it exists, and replication data is not needed. The estimator makes use of the empirical phase function, a normalized version of the empirical characteristic function. The empirical phase function was considered in the context of density deconvolution by \cite{delaigle2016} and \cite{nghiem2018density}. The method has two assumptions: the measurement error $U$ is symmetric around 0 with strictly positive characteristic function, and the distribution of $X$ is \textit{a}symmetric. These assumptions are fundamental for the identifiability of the phase function of $X$, which forms the basis of the estimation procedure. The assumptions are discussed in greater detail in Section 2.1; see also \cite{delaigle2016} for an in-depth discussion.

The remainder of this article is organized as below. In Section 2, we introduce the phase function-based estimator, develop its asymptotic properties, and establishes a connection to the method of moments approach. Section 3 considers some computational aspects relating to the estimator, including estimating standard errors in practice. Section 4 presents a simulation study to illustrate the performance of the phase function estimator and compare it with existing methods. Section 5 applies the phase function estimator to a real dataset, and Section 6 contains some concluding remarks.

\section{Phase Function Minimum Distance Estimation}
\subsection{Phase Function-Based Estimation}

Consider the simple linear errors-in-variables model with observed sample $(W_i,Y_i)$, $i=1,\ldots,n$ where
\begin{equation}
Y_i = \beta_0 + \beta_1 X_i + \varepsilon_i \quad\text{and}\quad W_i = X_i +U_i.
\label{linear_errors-in-variables}
\end{equation}
Here, the $X_i\in \mathbb{R}$ are independent and identically distributed with asymmetric density function $f_X$, the $U_i\in\mathbb{R}$ and $\varepsilon_i\in\mathbb{R}$ are  independent and identically distributed with respective density functions $f_U$ and $f_\varepsilon$, both symmetric about $0$ and having strictly positive characteristic functions. Furthermore,the $X_i$, $U_i$ and $\varepsilon_i$ are assumed mutually independent. It should be noted that the method developed here can also be used in the more general setting where each error term $U_i$ and $\varepsilon_i$ has a unique density function, say $f_{U,i}$ and $f_{\varepsilon,i}$, as long as these are all independent, symmetric about $0$, and have strictly positive characteristic functions. However, for simplicity of exposition the scenario with common error densities $f_U$ and $f_\varepsilon$ is presented. As to the assumed positivity of the characteristic functions, we note that many commonly used continuous distributions in the application of regression and measurement error satisfy this condition. This includes the Gaussian, Laplace, and Student's $t$ distributions. In general, the only symmetric distributions excluded are those defined on bounded intervals, such as the uniform. In the context of density deconvolution, \cite{delaigle2016} assumed that the random variable $X$ does not have a symmetric component, i.e. there is \textit{no} symmetric random variable $S$ for which $X$ can be decomposed as $X=X_0+S$ for arbitrary random variable $X_0$. In the present setting, this strict assumption is not required. More specifically, we only require that the covariate $X$ not be symmetric.

Now, let $\phi_X(t) = \mathrm{E}\left[\exp{(itX)}\right]$ denote the characteristic function of a random variable $X$.  The phase function of $X$ is then defined as the normalized characteristic function,
\begin{equation}
\rho_X(t) = \frac{\phi_X(t)}{|\phi_X(t)|},
\label{phase_func}
\end{equation}
where $|z| = (z\bar{z})^{1/2}$ is the complex norm with $\bar{z}$ denoting the complex conjugate of $z$. We now present our first result that establishes a relationship between the phase functions of $W$ and $Y$.

\begin{lemma}
	Consider univariate random variables $W=X+U$ and $Y=\beta_{0}+\beta_1 X + \varepsilon$. Assume that $X$ asymmetric with phase function $\rho_X(t)$,  and that $U$ and $\varepsilon$ are symmetric about $0$ with strictly positive characteristic functions. The phase function for $Y$ is then given by \[\rho_Y(t) = \exp\left(it\beta_0\right) \rho_X(\beta_1 t) = \exp\left(it\beta_0\right) \rho_W(\beta_1 t).\]
Hence, the phase function of $Y$ can be fully specified in terms of $\rho_W(t)$, the phase function of $W$, and parameters $(\beta_0,\beta_1)$. 
\end{lemma}

The above lemma follows follows simply by noting that $W$ and $Y$ have characteristic functions $\phi_W(t) = \phi_X(t) \phi_{U}(t)$ and $\phi_Y(t) = \exp\left(it\beta_0\right)\phi_X(\beta_1 t) \phi_{\varepsilon}(t)$. When evaluating the respective phase functions, the error components $\phi_{U}(t)$ and $\phi_{\varepsilon}(t)$ are canceled out in due to their assumed positivity. A full derivation of the relationship is given in Section A.1 of the Supplemental Material.

Empirical estimates of the phase functions of $W$ and $Y$ can be obtained from a random sample $(W_j,Y_j),\ j=1,\ldots,n$. Define
\[ \hat{\rho}_W (t) = \dfrac{\hat{\phi}_W(t)}{\vert \hat{\phi}_W(t) \vert } = \dfrac{\sum_{j=1}^{n} \exp(itW_j)}
{\left[{\sum_{j=1}^{n}\sum_{k=1}^{n} \exp\left\{it\left(W_j-W_k\right)\right\}}\right]^{1/2}} , \] with a similar definition holding for $\hat{\rho}_Y (t)$. The empirical phase functions can now be used to construct minimum distance estimators of the model parameters $(\beta_0,\beta_1)$. Define statistic
\begin{equation}
D(b_0, b_1)  = \int_{-\infty}^\infty \left| \hat{\rho}_Y(t) - \exp{(itb_0)}\hat{\rho}_W(b_1t)\right|^2 w(t) dt,
\label{eq:D1}
\end{equation}
where the weight function $w(t)$ is chosen to ensure that the integral is well-defined. The estimator $(\hat\beta_0,\hat{\beta}_1)$ is then computed as the global minimizer of the function $D(b_0, b_1)$.

The above idea can be easily extended to the case of multivariate regression with both error-prone and error-free covariates. Consider the model 
$Y= \beta_0 + X^\top\beta_1 + Z^\top\beta_2 + \varepsilon$ where $X, \beta_1 \in \mathbb{R}^{p_1}$ and $Z, \beta_2 \in \mathbb{R}^{p_2}$. Here, $Z$ represents the covariates measured without error. As before, let $W = X+U$ denote the contaminated version of $X$ where $U$ is $p_1$-dimensional symmetric measurement error.  Let $V = \beta_0 + X^\top\beta_1 + Z^\top\beta_2$ so that $Y=V+\varepsilon$. It then follows that $\rho_Y(t) = \rho_V(t)$.

Similarly, consider the linear combination in terms of the contaminated $W$, say \[\widetilde{V} = \beta_0 + W^\top \beta_1+Z^\top \beta_2 = V + U^\top \beta_1 = V + \widetilde{U}\] with $\widetilde{U}=U^\top \beta_1\in \mathbb{R}$ having distribution symmetric about zero with strictly positive characteristic function. It then also follows that $\rho_{\widetilde{V}}(t) =  \rho_V(t)$. Hence, the variables $Y$, $V$ and $\widetilde{V}$ all have the same phase function. To estimate $\beta=(\beta_0,\beta_1^\top,\beta_2^\top)^\top$, it is possible to construct a distance metric equivalent to (\ref{eq:D1}),
\begin{equation}
D (b_0,b_1,b_2) = \int_{-\infty}^{\infty} \left| \hat{\rho}_Y(t) - \exp{(itb_0)}\hat{\rho}_{\widetilde{V}}(t|b_1,b_2)\right|^2 w(t) dt
\label{eq:D_multiple}
\end{equation}
where, given $n$ random observations $(W_j , Z_j , Y_j)$, the empirical phase function corresponding to $\widetilde{V}$ is
\begin{equation}
\hat{\rho}_{\widetilde{V}}(t|\beta_1,\beta_2) = \dfrac{\sum_{j=1}^{n} \exp\left\{it\left(W_j^\top \beta_1 + Z_j^\top \beta_2\right)\right\}}{\left(\sum_{j=1}^{n}\sum_{k=1}^{n} \exp\left[it\left\{\left(W_j-W_k\right)^\top\beta_1+\left(Z_j-Z_k\right)^\top\beta_2\right\}\right]\right)^{1/2}}. 
\label{eq: multiple}
\end{equation}
Note that the statistic (\ref{eq:D_multiple}) does not treat the variables measured with and without error any differently. As such, the phase function approach could be implemented without knowledge of which variables are subject to measurement error. The estimate of $\beta=(\beta_0,\beta_1^\top,\beta_2^\top)^\top$ is found by minimizing $D(b_0,b_1,b_2)$.

\subsection{Asymptotic Properties of Phase Function Estimators}
\label{sec:as_prop}

In this section, we verify that the estimators obtained by minimizing statistic $D$ in (\ref{eq:D1}) satisfy the conditions required of M-estimators, and are therefore asymptotically normal. To this end, we first establish the almost sure convergence of $D$ to an appropriate limit. Note that, while the asymptotic properties of the phase function-based estimator are considered in the context of a simple linear errors-in-variables model, the results easily extend to the multivariate case. 

\begin{lemma}
	Assume that independent pairs $(W_1,Y_1),\ldots,(W_n,Y_n)$ are observed with $W_i=X_i+U_i$ and $Y_i=\beta_{0}+\beta_{1}X_i+\varepsilon_i$, with the distribution of $X_i$ asymmetric, and with $U_i$ and $\varepsilon_i$ having distributions symmetric about $0$ and with strictly positive characteristic functions. Furthermore, let $w(t)$ be a non-negative weight function with bounded support, taken without loss of generality to be $[-c,c]$.

	For this choice of weight function, the statistic $D(b_0 , b_1)$ defined in (\ref{eq:D1}) converges almost surely to a limit $D_\mathrm{true}(b_0,b_1)$ with 
	\[
	D_{\mathrm{true}}(b_0,b_1) = \int_{-\infty}^{\infty} \left|\rho_Y(t)-\exp(it b_0)\rho_W(b_1 t) \right|^2 w(t) dt. 
	\]
	The limit has unique global minimum $D_\mathrm{true}(\beta_0,\beta_1)=0$.
\end{lemma}

The proof of this lemma follows upon noting the empirical characteristic functions $\hat{\phi}_W(t)$ is an unbiased estimator of the true characteristic function $\phi_W(t)$ and converges almost surely to $\phi_W(t)$ on any bounded interval $[-c,c]$, see Theorem 2.1 of \cite{feuerverger1977empirical}. Applying the continuous mapping theorem, the empirical phase function $\hat{\rho}_W(t)$ also converges almost surely to the true phase function $\rho_W(t)$ on $[-c,c]$, and is an asymptotically unbiased estimator thereof. The convergence of $D$ to $D_\mathrm{true}$ follows from this. Next, noting that a phase function is uniquely identified by the asymmetric part of the corresponding distribution, the function $D_{\mathrm{true}}$ has a global minimum of $0$ at the true parameter values $(\beta_0,\beta_1)$.

\begin{theorem}
	Assume that conditions (i) and (ii) from Lemma 1 hold. Let $\hat{\beta}=(\hat{\beta_{0}},\hat{\beta_{1}})^\top$ denote the minimizer of $D$ in (\ref{eq:D1}). This estimator is consistent for the true $\beta=(\beta_0,\beta_1)^\top$, and is asymptotically normal,
	\begin{equation}
	\sqrt{n} \left(\hat{\beta} - \beta \right) \rightarrow \textit{N}\left(0, B^{-1}AB^{-1}\right)
	\end{equation}
	where
	\begin{equation}
	A = E\left(\lambda\lambda^{\top}\right)\ and\ B = E \left( \displaystyle\frac{\partial \lambda}{\partial \beta^\top} \right) 
	\label{A_B_matrices}
	\end{equation}
	with $\lambda = \displaystyle\frac{\partial D}{\partial \beta}$.
\end{theorem}

The consistency of $\hat{\beta}$ follows from Lemma 1 above and Theorem 5.7 in \cite{vandervaart2000asymptotic}. Having established consistency, and noting that $D$ has infinitely many bounded and continuous derivatives, asymptotic normality follows from Theorem 5.21 in \cite{vandervaart2000asymptotic}.

\subsection{Connection to Method of Moments Estimation}

\cite{delaigle2016} show that for any random variable $X$ with infinite number of moments, the phase function of $X$ can be expressed as
\[ \rho_X(t) =\exp\left\{\sum_{j=1}^{\infty} \dfrac{(-1)^{j+1} t^{2j-1}\kappa^{X}_{2j-1}}{(2j-1)!} \right\},
\]
where $\kappa_j^{X}$ denotes the $j$th cumulant of $X$. In other words, if the infinite series above converges, the phase function is determined uniquely by the odd-order cumulants of $X$. In this context, consider the model (\ref{linear_errors-in-variables}). If $X$, $U$, and $\varepsilon$ have an infinite number of finite moments,  the same holds true for $W$ and $Y$. Specifically, for $(W,Y)$ following the linear errors-in-variables model, it follows that
\begin{equation}
\exp\left\{i \sum_{j=1}^{\infty} \dfrac{(-1)^{j+1} t^{2j-1}\kappa^{Y}_{2j-1}}{(2j-1)!}\right\} = \exp\left[i \left\{t\beta_0 + \sum_{j=1}^{\infty} \dfrac{(-1)^{j+1} (\beta_1t)^{2j-1}\kappa^{W}_{2j-1}}{(2j-1)!}\right\}\right].
\label{phase_func_eq}
\end{equation}
One can use (\ref{phase_func_eq}) and match the coefficients of $t^{2j-1}$ to determine the relationship between the $j$th odd cumulants of $W$ and $Y$. For example, considering the coefficients of $t$ and $t^3$ gives $\kappa^Y_1  = \beta_0 + \beta_1\kappa^W_1$ and $\kappa^Y_3  = \beta_1^3 \kappa^W_3$.

Now, using properties of the complex norm, it follows that
\begin{equation*}
\frac{1}{4}\left|\rho_Y(t)-\exp(it \beta_0)\rho_W(\beta_1 t)\right|^2 = \sin^2\left\{\sum_{j=1}^{\infty} \dfrac{(-1)^{j+1}t^{2j-1} (\kappa^{Y}_{2j-1}-\beta_1^{2j-1}\kappa^{W}_{2j-1})}{(2j-1)!} - t\beta_0 \right\}.
\end{equation*}
When inference is based on the sample phase functions, the population cumulants above are replaced by their sample counterparts, and minimizing (\ref{eq:D1}) is equivalent to choosing the parameters $\beta_0$ and $\beta_1$ such that a function of the difference of all odd cumulants is minimized. As such, when the underlying distributions have an infinite number of moments, the phase function approach can be thought of as a method of moments-type approach that makes use of all \text{odd} cumulants of the variables of interest.

\section{Computational Considerations}
\label{sec:computing}
\subsection{Computing the Estimators}

Direct minimization of statistics \eqref{eq:D1} and \eqref{eq:D_multiple} is generally computationally expensive. In this section, a computational method is proposed that leads to faster calculation of the estimators. The idea is presented for the univariate errors-in-variables model, but can easily be extended to the multivariate model setting.  

\begin{lemma}
	Consider the statistics (\ref{eq:D1}) with weight function \[
	w(t) = K_{t^*}(t)  \left[\sum_{j=1}^{n}\cos(tY_j) \sum_{j=1}^{n} \cos\left\{t\left(b_0+b_1W_j\right)\right\}\right]^2
	\]
	where $K_{t^*}(t)=K(t/t^*)$ and $K(t)$ is a non-negative kernel function with bounded support on some interval $[-1,1]$. Minimization of (\ref{eq:D1}) is then equivalent to minimization of
	\begin{equation}
	D(b_0,b_1) =  \int_{0}^{t^*} \left[  \sum_{i=1}^n \sum_{j=1}^n \sin\left\{t(Y_i-b_0-b_1 W_j)\right\}\right]^2 K_{t^*}(t) dt 
	\label{eq:Dest2}
	\end{equation}
\end{lemma}

The derivation of this lemma is given in the Section A.2 of the  Supplementary Material. It is based on an application of Euler's formula, and noting that the ratio of the imaginary and real components of the phase function are equal to the same ratio calculated from the characteristic function. Formula \eqref{eq:Dest2} has computational complexity $\mathcal{O}(n^2)$. However, by some algebra it can be re-expressed as
\begin{equation}
D(b_0,b_1) =  \int_0^{t^*} \left[  S_y\sum_{j=1}^n \cos\left\{t(b_0+b_1 W_j)\right\} - C_y\sum_{j=1}^n \sin\left\{t(b_0+b_1 W_j)\right\} \right]^2 K_{t^*}(t) dt,
\label{eq:Dest}
\end{equation}
with $C_y = \sum_{j=1}^n \cos(tY_j)$ and $S_y=\sum_{j=1}^n \sin(tY_j)$. Evaluating \eqref{eq:Dest} has computational complexity $\mathcal{O}(n).$ The particular choice of weight function avoids instabilities that can occur in (\ref{eq:D1}) as a result of dividing by numbers close to $0$. With regards to choosing an appropriate constant $t^*$, we follow the suggestion in \cite{delaigle2016} who let $t^*$ be the smallest $t>0$ such that $ \vert \hat{\phi}_Y(t) \vert \leq n^{-1/4}$. 

When considering simplification of statistic (\ref{eq:Dest2}), It is also possible to eliminate the integral in the equation. To this end, let $\phi_{K,h}(\alpha) = \int_{-h}^{h} \cos(\alpha t) K_h(t) dt$. It then follows that
\begin{equation}
D(b_0,b_1) \propto \sum_{i,j,k,l} \Big[\phi_{K,t^*}\left\{Y_i - Y_k - b_1 (W_j-W_l)\right\} -\phi_{K,t^*}\left\{Y_i + Y_k + 2b_0 + b_1 (W_j+W_l)\right\} \Big].
\label{eq:Quadsum}
\end{equation}
Note that while expression (\ref{eq:Quadsum}) eliminates the need to numerically evaluate an integral as in (\ref{eq:Dest}), we generally found that the form in (\ref{eq:Dest}) was much faster to compute than the expression involving the quadruple sum in (\ref{eq:Quadsum}).

Now, considering again the recommended computational form in (\ref{eq:Dest}). By an application of the Leibniz rule, the first partial derivatives of $D$ with respect to $b_0$ and $b_1$, denoted here $\lambda(b_0,b_1) = \left\{\lambda_0(b_0,b_1), \lambda_1(b_0,b_1) \right\}^\top$, are
\begin{equation}
\lambda_0 = \frac{\partial D}{\partial b_0} = -2 \int_0^{t^*} \left[  \sum_{i,j} \sin\left\{t(Y_i-b_0-b_1 W_j)\right\}\right]\left[  \sum_{i,j} \cos\left\{t(Y_i-b_0-b_1 W_j)\right\}\right] K_{t^*}(t) dt 
\label{partial1}
\end{equation}
and
\begin{equation}
\lambda_1 = \frac{\partial D}{\partial b_1} = -2 \int_0^{t^*} \left[  \sum_{i,j} \sin\left\{t(Y_i-b_0-b_1 W_j)\right\}\right] \left[ \sum_{i,j} W_j \cos\left\{t(Y_i-b_0-b_1 W_j)\right\}\right] t K_{t^*}(t) dt .
\label{partial2}
\end{equation}
The expressions for $\lambda_0$ and $\lambda_1$ can be used as estimating equations to solve for $(\hat{\beta_{0}},\hat{\beta_{1}})$. These expressions will also be useful in the next section when considering the estimation of standard errors for the estimators.

\subsection{Standard Error Estimation}
\label{sec:se_estim}

We now consider estimation of the covariance matrix of $\hat{\beta}$.  Using the asymptotic variance as given in Theorem 1 would be reasonable, but direct evaluation of matrices $A$ and $B$ in (\ref{A_B_matrices}) is not possible as this requires knowledge of the distributions of $X$, $U$ and $\varepsilon$. If these distributions were known, a likelihood approach could be used for parameter estimation rather than the proposed phase function approach. 

The bootstrap is a A popular method for estimating the covariance matrix of estimated parameters in a nonparametric setting such as this is the bootstrap. This requires repeated calculation of bootstrap estimators $\hat{\beta}^*_b$ based on bootstrap samples $(W_{b,i}^*,Y_{b,i}^*)$, $i=1,\ldots,n$ for $b=1,\ldots,B$ drawn with replacement from the observed sample. The estimated covariance matrix is then taken to be the sample covariance matrix of the bootstrap replicates $\hat{\beta}^*_b$. The procedure can be slow due to the repeated evaluation of a computationally expensive loss function for each bootstrap sample. Implementation is described in Algorithm 1.

We propose here a modified bootstrap algorithm for estimating the standard errors that combines bootstrap methodology with approximation of matrices $A$ and $B$ in \eqref{A_B_matrices}. To this end, note that matrix $A$ is the covariance matrix of $\lambda$, the first partial derivatives of \eqref{eq:Dest2}  given by \eqref{partial1} and \eqref{partial2} in the univariate setting. As such, bootstrap methodology can be used to estimate matrix $A$, while $B$ can estimated by evaluating the second derivatives of $D$ at the parameter estimates $\hat{\beta_{0}}$ and $\hat{\beta_{1}}$. Expressions for these second derivatives are unwieldy, but are easily evaluated numerically; see Section A.3 in the supplemental material. We refer to this approach as the \textit{plug-in bootstrap} approach and outline implementation in Algorithm 2. Note that the plug-in covariance matrix is orders of magnitude faster to compute that the boostrap estimator as it does not require repeated minimization of a statistic involving numerical integration.

\begin{algorithm}
	\caption{Full Bootstrap Algorithm}
	\begin{itemize}
		\item For $b = 1, \dots , B$
		\begin{itemize}
			\item Sample $n$ pairs with replacement from the observed data to obtain bootstrap sample $(W^*_{i,b}, Y^*_{i,b})$, $i=1,\ldots,n$. 
			\item Calculate the bootstrap estimators $\hat{\beta}_b^*$ by minimizing (\ref{eq:Dest2}) using the bootstrap sample.
		\end{itemize}
		\item Calculate the empirical covariance matrix of the bootstrap statistics $\hat{\beta}_1^*,\ldots,\hat{\beta}_B^*$,
		\begin{equation}
		\hat{\boldsymbol{\Sigma}}_{\mathrm{boot}} = \frac{1}{B}\sum_{b=1}^{B} \left(\hat{\beta}_b^*-\bar{\beta}^*\right)\left(\hat{\beta}_b^*-\bar{\beta}^*\right)^\top
		\label{boot_cov}
		\end{equation}
		where $\bar{\beta}^* = B^{-1}\sum_{b} \hat{\beta}_b^*$ is the mean of the bootstrap replicates.
	\end{itemize}
\end{algorithm}

\begin{algorithm}
	\caption{Plug-in Bootstrap Algorithm}
	\begin{itemize}
		\item For $b = 1, \dots , B$
		\begin{itemize}
			\item Sample $n$ pairs with replacement from the observed data to obtain bootstrap sample $(W^*_{i,b}, Y^*_{i,b})$, $i=1,\ldots,n$. 
			\item Calculate $\lambda^*_b=\lambda^*_b(\hat{\beta}_{0},\hat{\beta}_{1})$ as in (\ref{partial1}) and (\ref{partial2}) using the $b$th bootstrap sample.
		\end{itemize}
		\item Calculate 
		\begin{equation*}
		\hat{A}_{\mathrm{boot}} = \frac{1}{B}\sum_{b=1}^{B} {\lambda}_b^* {\lambda}_b^{*\top} \quad \mathrm{and}\quad \hat{B} = \left[\frac{\partial \lambda}{\partial [b_0,b_1]^\top} \right]_{(b_0,b_1)=(\hat{\beta}_{0},\hat{\beta}_{1})}. 
		\end{equation*}
		\item Calculate plug-in covariance matrix $\hat{\Sigma}_{\mathrm{plug}} = \hat{B}^{-1} \hat{A}_{\mathrm{boot}} \hat{B}^{-1}$.
	\end{itemize}
\end{algorithm}

\section{Simulation Study}

An extensive simulation study was conducted to evaluate the performance of the phase function-based estimators for various underlying distributions. In this section, we report and discuss a representative selection of these simulation results.

First, parameter estimation was explored in the univariate setting. Data were generated according to the model
$Y_i = \beta_0 + \beta_1 X_i + \varepsilon_i$ and $W_i = X_i + U_i$, $i= 1,\ldots, n$ with true parameters $(\beta_0,\beta_1) = (1,3)$. Three asymmetric distributions were used to simulate $X$, namely (1) a half-normal distribution, $X \sim |N(0,1)|$, (2) an exponential distribution, $X \sim \exp(1)$, and (3) a bimodal mixture distribution, $X \sim 0.5N(5,1^2)+0.5N(2.5,0.6^2)$. Three different distributions were considered for error components $U$ and $\varepsilon$, namely the normal, $t$-distribution with $2.5$ degrees of freedom, and the Cauchy distribution. For the Normal and $t_{2.5}$ distributions, the error components were simulated to have mean $0$ and respective variances $\sigma_U^2$ and $\sigma_\varepsilon^2$. For the Cauchy distribution, the error components were simulated to be symmetric about $0$ and have respective interquartile range (IQR) $\sigma_U$ and $\sigma_\varepsilon$. The variance and IQR parameters were chosen to achieve specific noise-to-signal ratios, $p_W = \sigma_U^2 / \sigma_X^2$ and $p_Y = \sigma_{\varepsilon}^2 / (\beta_1 \sigma_X)^2$. The noise-to-signal ratios pairs reported here are $ (p_W,p_Y)=(0.25, 0.40)$. Results are reported for sample sizes $n \in \{500, 1000\}$. Simulation with other noise-to-signal ratios were carried out, and these results are reported in the Section C.2 of the Supplement Material. For each possible configuration of simulation specifications, $N=2000$ samples were generated. 

For the Normal and $t_{2.5}$ error cases, four different estimators were calculated for each simulated dataset. First, the naive estimators ignoring measurement error were obtained by regressing the contaminated $W$ on $Y$. Second, the generalized method of moments estimators using $M=3$ moments were computed. Three different choices of weight function were considered for the phase function estimator. Table C.1 in the supplemental material compares the resulting estimators. As the weight function $K(t) = (1-|t|)^2\mathbb{I}(|t|\leq1)$ was found to have consistently good performance,the corresponding results are reported here. Finally, the disattenuated regression estimators were also calculated. For disattenuation, the parameters $(\sigma^2_U,\sigma^2_X)$ were treated as known quantities, and would not be computable in practice under the minimal model assumptions for the phase function method. For the Cauchy error case with infinite variance, no analog for disattenuation is known. However, even though there is no theoretical justification for doing so, the generalized method of moments estimators were computed to compare to the phase function estimators.

Now, letting $\hat{\beta}_{m,j}^{(\mathrm{method})} $ denote the estimator of $\beta_j$ calculated for the $m$th sample with the superscript ``method'' a placeholder for a specific method from those listed above, define squared error $\mathrm{SE}_{m,j}^{(\mathrm{method})} = [\hat{\beta}_{m,j}^{(\mathrm{method})} - \beta_j]^2$. As both the generalized method of moments and phase function estimators are very prone to outliers in small samples, the median square errors is used rather than mean square error, as the former is more robust against these outliers. For the Normal and $t_{2.5}$ error case, we report in Table \ref{tab:sim_Uerrors-in-variables} the \textit{ratios} of median square errors for the naive, generalized method of moments, and phase function estimators relative to the disattenuated estimators. An entry in the table larger than $1$ indicates superior performance of the disattenuated estimators, while an entry smaller than $1$ indicates superior performance of the associated method. Entries can also be compared across methods, with a larger entry indicating worse performance of a method for a given set of simulation specifications. The full simulation results, including the median squared error and a robust estimate of standard error based in the interquartile range, are given in the Section C.2 of the Supplement Material.

\begin{table}
	\centering
	\small
	\caption{Ratio of median square error of estimators relative to the disattenuated regression estimators in the univariate model simulation with model errors being Normal and $t_{2.5}$ distributions. Note GMM stands for generalized method of moments.}
	\begin{tabular}{cc Hc | cc cc cc } \hline
		Error type & True $X$  & $(p_W, p_Y)$  & $n$  & \multicolumn{2}{c}{Naive}   & \multicolumn{2}{c}{GMM} & \multicolumn{2}{c}{Phase}     \\
		\cline{5-10}
		&  &  & & $\beta_0$ & $\beta_1$ & $\beta_0$ & $\beta_1$ & $\beta_0$ & $\beta_1$ \\
		\hline
		Normal &$|N(0,1)|$ & $(0.25, 0.40)$ & 500 &  73.0 & 98.7 & 2.35 & 2.81 & \bf{2.15} & \bf{2.42} \\   
		&& & 1000  & 131.3 & 189.8 & 2.55 & 3.16 & \bf{1.83} & \bf{2.32} \\ 
		&$\mathrm{exp}(1)$& $(0.25,0.40)$ & 500 & 44.3 & 67.5 & \bf{1.18} & \bf{1.15} & \bf{1.18} & 1.59 \\  
		&& & 1000 & 89.0 & 129.4 & \bf{1.27} & \bf{1.24} & 1.36 & 1.77 \\ 
		&Bimodal& $(0.25,0.40)$ & 500 & 100.7 & 118.4 & 10.1 & 11.5 & \bf{5.71} & \bf{6.48} \\ 
		&& & 1000  & 200.2 & 235.0 & 9.74 & 11.2 & \bf{4.02} & \bf{4.74} \\  
		\hline
		$t_{2.5}$& $|N(0,1)|$& $(0.25, 0.40)$ & 500 & 5.89 & 6.18 & 0.30 & 0.29 & \bf{0.16} & \bf{0.18}  \\   
		&& & 1000  & 8.75 & 9.32 & 0.29 & 0.29 & \bf{0.09} & \bf{0.12}  \\ 
		&$\mathrm{exp}(1)$& $(0.25,0.40)$ & 500 & 6.64 & 5.88 & 0.17 &\bf{0.12} & \bf{0.14} & 0.19 \\  
		&& & 1000 & 9.09 & 8.75 & 0.16 & \bf{0.11} & \bf{0.10} & 0.12  \\ 
		&Bimodal& $(0.25,0.40)$ & 500 & 6.29 & 6.10 & 1.26 & 1.21 & \bf{0.27} & \bf{0.24} \\ 
		&& & 1000  & 9.29 & 9.12 & 1.46 & 1.42 & \bf{0.12} & \bf{0.11}\\  
		\hline
	\end{tabular}
	\label{tab:sim_Uerrors-in-variables}
\end{table}

Considering the results in Table \ref{tab:sim_Uerrors-in-variables}, we note that the naive estimator performs the worst among all the considered estimators across all simulation configurations. This is to be expected due to the known bias when not correcting for measurement error. For normally distributed errors, the phase function estimator performs better than generalized method of moments for both the cases $X$ distributed as half-normal and as a bimodal mixture of normals. The improvement of the phase method is especially dramatic in the bimodal $X$ case considered. On the other hand, for $X$ having an exponential distribution, generalized method of moments performs better than the phase function method. 

We reach similar conclusions when considering the case of a $t_{2.5}$ distribution for the error. Overall, the phase function method has superior performance for the cases $X$ half-normal and $X$ bimodal. In the case of $X$ having an exponential distribution, generalized method of moments does better at estimating the slope $\beta_1$, while the phase function method does better at estimating the intercept $\beta_{0}$. We initially found the good performance of generalized method of moments surprising, as its implementation here makes use of the third sample moments, whereas third moments do not exist for the error distribution used. However, generalized method of moments downweights the second and third sample moments using the fourth through sixth sample moments. Intuitively, the latter quantities will be really large, to some extent regulating the effect of using the former on estimating parameters. Still, the performance of the phase function method is generally far superior in this setting. In fact, noting that most of the median square error ratios are much smaller than $1$, the phase function method is seen to be superior to correcting for attenuation using known error variances.

\begin{table}[ht]
	\centering
	\caption{Median square error of the generalized method of moments estimators, denoted GMM in the table, and the phase function estimators when the model errors are Cauchy.}
	\begin{tabular}{cc|cccc}
		\hline
		True $X$ & $n$ & \multicolumn{2}{c}{GMM}& \multicolumn{2}{c}{Phase} \\ 
		\cline{3-6}
		&& $\beta_0$ & $\beta_1$ & $\beta_0$ & $\beta_1$\\
		\hline
		$\vert N(0,1) \vert$  & 500 & 4.44 & 8.99 & 0.05 & 0.10 \\ 
		& 1000 & 4.42 & 9.00 & 0.02 & 0.04 \\ 
		$\exp(1)$ & 500 & 5.43 & 8.99 & 0.03 & 0.05 \\ 
		& 1000 & 6.29 & 9.00 & 0.01 & 0.03 \\ 
		Bimodal & 500 & 52.39 & 8.94 & 2.32 & 0.15 \\ 
		& 1000 & 53.60 & 8.98 & 1.85 & 0.12 \\ 
		\hline
	\end{tabular}
	\label{tab:Cauchy}
\end{table}
Table \ref{tab:Cauchy} presents the simulation results for the generalized method of moments and the phase function estimators when the model errors follow a Cauchy distribution. In all the considered settings, the phase function estimator has a much smaller median square error than the generalized method of moments. The poor performance of generalized method of moments is expected because no moments exists for the Cauchy distribution. The phase function method, however, still performs well as it does not rely on the existence of error moments.

A second simulation study was done using two predictors, one measured with error and one without. Data were simulated according to the model $Y_i = \beta_0 + \beta_X X_i + \beta_Z Z_i + \epsilon_i$, $W_i = X_i + U_i$, $i=1,\ldots, n$ with parameters $\beta_0 = 0, \beta_X = 3$, and $\beta_Z = 2$. Here, $X$ is the error-prone covariate while $Z$ is error-free. Samples sizes $n \in \{1000, 2000\}$ were considered. We include here results for the two cases $X$ half-normal and $X$ having the bimodal normal mixture defined at the start of this section. The covariate $Z$ was generated from the same distribution as $X$, and a normal copula with $\rho = 0.5$ was used to generate $X$ and $Z$ to be correlated. The error distributions were taken to be normal with noise-to-signal ratios as in the univariate simulation. For each simulation configuration, 2000 replications were run. For each run, the phase function estimators and the naive estimators for both $\beta_X$ and $\beta_Z$ were computed. Furthermore, simulation-extrapolation  of \cite{stefanski1995simulation} was also implemented using the known measurement error variance. When the measurement error variance is unknown or not estimable, simulation-extrapolation  cannot be used. It is therefore included for comparative purposes. Table \ref{tab:mse_multtiple} reports again the ratio of median square error for the naive and phase function methods relative to the simulation-extrapolation  estimators. As before, see Section C.3 of the supplement material for the full simulation results.

\begin{table}[h!]
	\centering
	\caption{Ratio of median square error of estimators relative to the simulation-extrapolation  regression estimators in the bivariate model simulation.}
	\begin{tabular}{cHc|rrrr}
		\hline
		True $X$  & $(p_W, p_Y)$        & $n$    & \multicolumn{2}{c}{Naive} & \multicolumn{2}{c}{Phase}                         \\ 
		\cline{4-7}
		& & & $\beta_X$ & $\beta_Z$ & $\beta_X$ & $\beta_Z$ \\
		\hline
		$\vert N(0,1)\vert$& (0.25,0.40)  & 1000 & 107.1 & 35.4 & 15.0 & 40.2 \\ 
		&  & 2000 & 219.1 & 81.2 & 2.19 & 6.47 \\ 
		Bimodal & (0.25, 0.40)  & 1000 & 152.8 & 51.9 & 10.1 & 24.5 \\ 
		& & 2000 & 343.3 & 104.7 & 9.02 & 19.9 \\ \hline
	\end{tabular}
	\label{tab:mse_multtiple}
\end{table}

Again, the poor performance of the naive method in Table \ref{tab:mse_multtiple} is not surprising. The phase function method holds up well against simulation-extrapolation . It is clear that the method improves (in a relative sense) as the sample size increases from $1000$ to $2000$. Furthermore, the phase function approach has large relative median squared errors when $(p_W,p_Y)=(0.25,0.4)$, corresponding to large measurement error contamination. However, these scenarios also improve, sometimes dramatically so, when the same size increases.

Finally, we performed a simulation study to examine the performance of the (full) bootstrap and plug-in bootstrap methods for estimating standard errors of the parameters. Data were simulated from the univariate model used to generate Table \ref{tab:sim_Uerrors-in-variables}. For each simulated sample, both bootstrap methods were used to estimate the standard errors of the model coefficients. Reported here are the results for $X$ half-normal and $X$ bimodal, and two sample sizes $ n\in \{1000,2000\}$. For each simulation configuration, 2000 samples were generated. For each, the phase function estimates were computed. A total of $B=100$ bootstrap samples were generated for each of the methods described in Section \ref{sec:se_estim} to estimate standard errors. The true standard error was also estimated using the $1000$ pairs $\hat{\beta_{0}},\hat{\beta_{1}}$ estimated from the simulated data using the phase function methods. The median of $\sqrt{n}\times \widehat{\mathrm{se}}$, with $\widehat{\mathrm{se}}$ denoting estimated standard error, is reported in Table \ref{tab:var_errors-in-variables} for each method.

We note in Table \ref{tab:var_errors-in-variables} that the full bootstrap generally gives estimated standard errors very close to the true (Monte Carlo) values. The plug-in method has a tendency to over-estimate the standard error, especially for sample size $1000$. However, the plug-in method is superior in terms of computation speed. These computational time comparisons are based on running simulations on a distributed computing system with 80 nodes consisting of 36 cores each with 256GB memory and with an Intel Xeon E5-2695 v4 CPU. For sample size $n = 1000$, the average computation time for the full bootstrap around 34 minutes, while the plug-in bootstrap had an average computation time of around 1 minute. Similarly, for sample size $n = 2000$ the full and plug-in average computation times were around 49 minutes and 2 minutes, respectively. In many instances, one might be willing to use a method that over-estimates the size of the standard error for this type of speed-up in computation.
\newcolumntype{H}{>{\setbox0=\hbox\bgroup}c<{\egroup}@{}}
\begin{table}[ht]
	\centering
	\caption{True standard error (Monte Carlo) and median of estimated standard error, scaled by the square root of the sample size, using two different bootstrap approaches.}
	\begin{tabular}{cHc | c cccc c}
		\hline
		True $X$ & (NSR) & $n$            & \multicolumn{2}{c}{Monte Carlo} & \multicolumn{2}{c}{Bootstrap$_\mathrm{full}$} & \multicolumn{2}{c}{Bootstrap$_\mathrm{plug}$}  \\
		\cline{4-9}
		& & & $\beta_0$ & $\beta_1$ & $\beta_0$ & $\beta_1$ & $\beta_0$ & $\beta_1$ \\
		\hline
		$|N(0,1)|$&$(0.25, 0.40)$ & $1000$  & 0.48 & 0.56 & 0.48 & 0.55 & 0.56 & 0.71  \\ 
		& & $2000$  & 0.39 & 0.46 & 0.40 & 0.46 & 0.42 & 0.53 \\ 
		Bimodal& $(0.25, 0.40)$ & $1000$  & 2.82 & 0.75 & 2.95 & 0.79 & 5.27 & 1.36  \\
		& & $2000$  & 2.26 & 0.60 & 2.22 & 0.59 & 2.90 & 0.75      \\ 
		\hline\end{tabular}
	\label{tab:var_errors-in-variables}
\end{table}

\section{Air Quality Data Examples}
Here, we consider a dataset analyzed by \cite{de2008field} considering the measurement of carbon monoxide (CO) levels in present in an urban environment over time. The dataset is publicly available online at the UCI Machine Learning Repository (https://archive.ics.uci.edu/ml/datasets.html) and is labeled \textit{Air Quality}. In the experiment reported, a low-cost gas multi-censor device, also known as an electronic nose, was used to monitor atmospheric pollutants in an urban environment. Carbon monoxide was one of the pollutants being monitored and is of primary interest in our analysis. Measurements obtained by electronic noses use tin oxide as a proxy for carbon monoxide. These devices are also subject to measurement error, especially when compared to networks of spatially distributed fixed stations using industrial spectromoters. The latter are commonly used to monitor air pollution in urban environments, but use is restricted by cost and size considerations. The sources of measurement error for electronic noses range from known device stability issues to local atmospheric dynamics. Even so, it is desirable to consider the proper calibration of electronic noses for supplemental use in monitoring air pollution in urban areas. Specifically, we consider estimating the true relationship between tin oxide (subject to measurement error) and carbon monoxide.

\begin{figure}[!t]
	\centering
	\includegraphics[width=14cm, trim = {0 0cm 0 0cm}, clip]{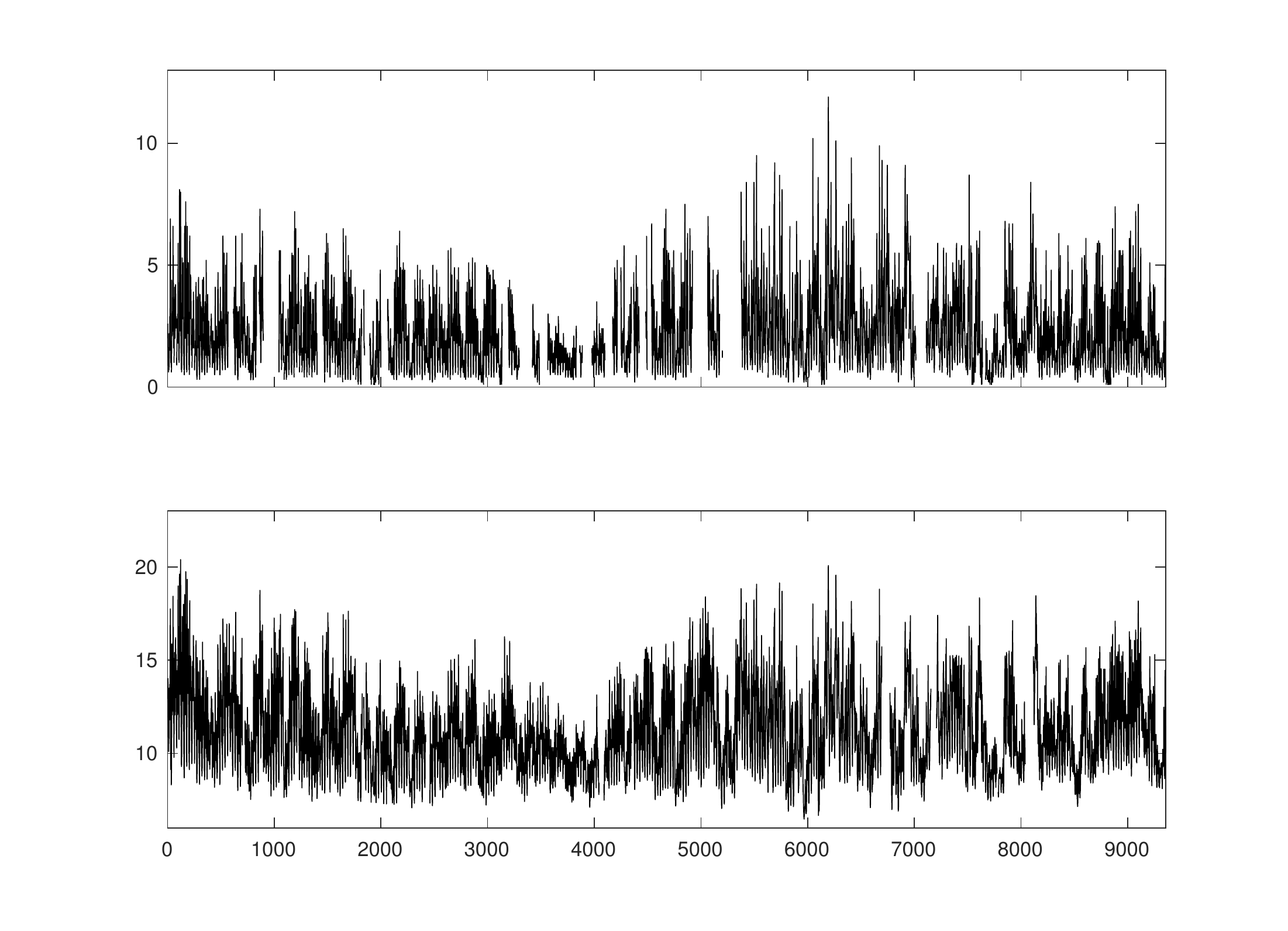}
	\caption{Time series plots for carbon monoxide $Y_t$ (top)  and the average sensor output $W_t$ (bottom).}
	\label{fig:airQuality}
\end{figure}

The experiment, which lasted 13 months, was performed at a main road with heavy traffic in an Italian city. During this period, hourly observations were collected from both an electronic nose ($W$ data) and a distributed network of seven fixed stations ($Y$ data). The measurements represent hourly averages of data collected at $8$ second increments. The data are denoted $(W_{t},Y_{t}),\ t=1,\ldots,T$, with $T=9357$ hourly periods transpiring during the experiment. However, $2013$ of these have partially or completely missing data, leaving $7344$ complete observations for the analysis. Time series plots of the measurements are shown in Figure \ref{fig:airQuality}. Note that the $W$ measurements in the figure and throughout the analysis are equal to the original data divided by $100$.

To account for time-of-day effects on pollution levels, the data were de-trended. To this end, let $\mathcal{I}_k = \{t: t = k+24(j-1),\ j = 1,2,\ldots\ \mathrm{and}\ t\leq T \}$ with $k=1,\ldots,24$ denote the collection of indices corresponding to measurements at hour $k$. Define observed hourly mean $\hat{\mu}_k = |\mathcal{I}_k|^{-1} \sum_{t\in \mathcal{I}_k} W_t$ for $k=1,\ldots,24$. The expression for $\hat{\mu}_k$ makes use of a slight abuse of notation, as the sum is only taken over indices corresponding to non-missing observations. The de-trended data are calculated as \[\tilde{W}_{t} = W_t - \sum_{k=1}^{24}\mathbb{I}\{t\in \mathcal{I}_k \}\hat{\mu}_k,\ t=1,\ldots,T. \]
The de-trended $\tilde{Y}_t$ are defined in an analogous manner, resulting in pairs $(\tilde{W}_t,\tilde{Y}_t)$, $t=1,\ldots,T$. It is now assumed that $\tilde{Y}_{t} = \beta_1 X_{t} + \varepsilon_{t}$ and $\tilde{W}_{t} = X_{t} + U_{t}$ with $X_t$ denoting the true CO level at time $t$. Note the lack of intercept term $\beta_0$ in the model. This is a result of de-trending the data. We assume that the error components $U_t$ and $\varepsilon_t$ are independent, and that these error components are independent of stationary time series $X_t$. All variables are assumed to have finite variance. The stationarity of $X_t$ is important as this ensures that the empirical phase functions still a consistent estimate $\rho_X$.

The generalized method of moments and phase function estimators of slope $\beta_1$ were calculated. To account for the correlation structure in $X_t$, the block bootstrap with block length $L=192$ was used to estimate the associated standard errors, see \cite{kunsch1989jackknife} for details on this technique. The generalized method of moments estimator is $\hat{\beta_{1}}^{(\mathrm{\mathrm{GMM}})}=0.73$ with estimated standard error $0.07$. The phase function estimator is $\hat{\beta_{1}}^{(\mathrm{phase})}=0.71$ with estimated standard error $0.02$. The naive estimator of slope is $0.52$, indicating the strong effect of measurement error here. Comparing the phase function and naive estimators of slope using the known attenuation relationship suggests the proportion of error variance is $0.36$. Moreover, the generalized method of moments and phase function estimates seemingly correct for the exogenous contamination present in the electronic nose measurements. While the two estimators are close to one another, the standard error of generalized method of moments is substantially larger than that of the phase function estimator.

\section{Conclusion}

The proposed phase function methodology is a new solution to the linear errors-in-variables problem where replicate data and/or prior knowledge of measurement error variance are not available.  Contamination of the observed features should not be ignored when making an inference, but strong model requirements can make it difficult to appropriately correct the error and leave the practitioner with a biased model. To our knowledge, the only solution not making such strict assumptions is the generalized method of moments. Our proposed method is seen to be competitive with generalized method of moments, and often has much smaller median squared error. Furthermore, the phase function-based method does not rely on the existence of an arbitrary number of moments. Future work will consider combining the strengths of the generalized method of moments and phase function methods: generalized method of moments can be implemented when the underlying variable has a symmetric distribution, whereas the phase function method requires asymmetry of the underlying distribution.

\bibliographystyle{biometrika}
\bibliography{citations}

\newpage
\section*{Supplement Material}
\makeatletter
\renewcommand{\thetable}{\thesection.\@arabic\c@table}
\@addtoreset{table}{section}
\makeatother
\renewcommand\thesection{\Alph{section}}
\setcounter{section}{0}
\section{Technical Results}
\subsection{Proof of Lemma 1}
Recall that $W = X + U$ with $X$ is independent of $U$. By independence, the characteristic function of $W$ is given by
\[
\phi_W(t) = \mathbb{E}(e^{itW}) = \mathbb{E}(e^{it(X + U)}) = \mathbb{E}(e^{itX})\mathbb{E}(e^{itU}) = \phi_X(t) \phi_U(t).
\]
By assumption, the characteristic function of $U$ satisfies $\phi_U(t) = \vert \phi_U(t) \vert$ for $t$.  Thus, the phase function for $W$ is 
\[
\rho_W(t) = \frac{\phi_W(t)}{\vert \phi_W(t) \vert} = \frac{\phi_X(t) \phi_U(t)}{\vert \phi_X(t) \vert \vert \phi_U(t) \vert} = \frac{\phi_X(t)}{\vert \phi_X(t) \vert} = \rho_X(t).
\]
Subsequently, the random variables $W$ and $X$ have the same phase function.  

Similarly, random variable $Y = \beta_0 + \beta_1 X + \varepsilon$ has characteristic function
\[
\phi_Y(t) = e^{it\beta_0}\phi_X(t\beta_1)\phi_\epsilon(t).
\]
with phase function given by
\begin{equation}
\rho_Y(t) = \frac{e^{it\beta_0}\phi_X(t\beta_1)\phi_\epsilon(t)}{\vert e^{it\beta_0} \vert \vert \phi_X(t\beta_1)\vert \vert \phi_\epsilon(t) \vert} = 
e^{it\beta_0} \rho_X(t\beta_1) = e^{it\beta_0} \rho_W(t\beta_1),
\label{eq:phYW}
\end{equation}
establishing the relationship between the phase functions of $W$ and $Y$.

\subsection{Proof of Lemma 3}
For any complex number $z$, let $R(z) = \text{Im}(z)/\text{Re}(z)$ denote the ratio of the imaginary and real parts of $z$. Now, consider the relationship that exists between the phase functions of $Y$ and $W$ as given in \eqref{eq:phYW}, and recall that any phase function has norm equal to $1$ for all $t$. It follows that \eqref{eq:phYW} is equivalent to 
\begin{equation}
R[\rho_Y(t)] = R [\exp(it\beta_0)] R [\rho_W(\beta_1t)]. 
\label{eq:ratio}
\end{equation}
Furthermore, as the phase function is a scaled version of the characteristic function, $R[\rho_Y(t)] = R[\phi_Y(t)] $ and \eqref{eq:ratio} is equivalent to
\[
R[\phi_Y(t)] = R [\exp(it\beta_0)] R [\phi_W(\beta_1t)]. 
\]
By Euler's formula, this can be written as
\begin{equation}
\frac{E\left[\sin \left( tY \right)\right]}{E\left[\cos \left( tY \right)\right]} = \frac{E\left[\sin \left(t\left( \beta_0 + \beta_1 W\right) \right)\right]}{E\left[\cos \left(t \left( \beta_0 + \beta_1W\right) \right)\right]}. 
\end{equation}
Therefore, minimizing the statistics D in equation (3) of the main paper is equivalent to minimizing
\begin{equation}
D(b_0, b_1) = \int_{-\infty}^{\infty} \left( \dfrac{\sum_{j=1}^{n} \sin(tY_j)}{\sum_{j=1}^{n} \cos(tY_j)} - \dfrac{\sum_{j=1}^{n} \sin\left(t\left(b_0+b_1W_j\right)\right)}{ \sum_{j=1}^{n} \cos\left(t\left(b_0+b_1W_j\right)\right)} \right)^2 w(t) dt.
\label{eq:D2}
\end{equation}
If choosing the weight function as stated in the lemma, the integrand is an even function with respect to $t$. Then the result follows from simplifying the resulting trigonometric products. 

\subsection{Expressions for the second derivative of $D(b_0,b_1)$}

In Section 3.2 of the main paper, a plug-in bootstrap method is proposed for estimating the standard errors of the phase function estimators. Evaluation there requires calculation of the second derivatives of the distance metric $D$ evaluated at the estimators $\hat{\beta}_0$ and $\hat{\beta}_1$. These functions are reported here.

Specifically,

\[
\begin{split}
\dfrac{\partial^2 D}{\partial b_0^2 } = \sum_i \sum_j \sum_k \sum_l 
\int_0^{t^*} 2t^2 K_{t^*}(t)
\Bigg[ & \cos\left\{ t (Y_i - b_0 - b_1 W_j ) \right\}  \cos\left\{ t (Y_l - b_0 - b_1 W_k ) \right\} \quad \\
& - \sin\left\{ t (Y_i - b_0 - b_1 W_j ) \right\}  \sin\left\{ t (Y_l - b_0 - b_1 W_k ) \right\} \Bigg] dt,
\end{split}
\]

\[
\begin{split}
\dfrac{\partial^2 D}{\partial b_0 \partial b_1} = \sum_i \sum_j \sum_k \sum_l 
\int_0^{t^*} 2t^2 K_{t^*}(t)
\Bigg[ & W_j\cos\left\{ t (Y_i - b_0 - b_1 W_j ) \right\}  \cos\left\{ t (Y_l - b_0 - b_1 W_k ) \right\}  \quad \\
& - W_k \sin\left\{ t (Y_i - b_0 - b_1 W_j ) \right\}  \sin\left\{ t (Y_l - b_0 - b_1 W_k ) \right\}  \Bigg] dt,
\end{split}
\]
and  
\[
\begin{split}
\dfrac{\partial^2 D}{\partial b_1^2 } = \sum_i \sum_j \sum_k \sum_l 
\int_0^{t^*} 2t^2 K_{t^*}(t)
\Bigg[ & W_jW_k\cos\left\{ t (Y_i - b_0 - b_1 W_j ) \right\}  \cos\left\{ t (Y_l - b_0 - b_1 W_k ) \right\} \quad \\
& - W_k^2\sin\left\{ t (Y_i - b_0 - b_1 W_j ) \right\}  \sin\left\{ t (Y_l - b_0 - b_1 W_k ) \right\} \Bigg] dt. 
\end{split}
\]

The quadruple sums can be eliminated using some simple but tedious algebra, giving expressions that are computationally convenient,

\[
\begin{split}
\dfrac{\partial^2 D}{\partial b_0^2} =& 
\int_0^{t^*} 2t^2 K_{t^*}(t) 
\left[\sum_i \sum_j \cos\left\{t(Y_i - b_0 - b_1 W_j) \right\}\right]^2 dt \\
& - \int_0^{t^*} 2t^2 K_{t^*}(t) \left[\sum_i \sum_j \sin\left\{t(Y_i - b_0 - b_1 W_j) \right\}\right]^2 dt , 
\end{split}
\]

\[
\begin{split}
\dfrac{\partial^2 D}{\partial b_0 \partial b_1 } =& 
\int_0^{t^*} 2t^2 K_{t^*}(t) 
\left[\sum_i \sum_j \cos\left\{t(Y_i - b_0 - b_1 W_j) \right\}\right]\left[\sum_i \sum_j W_j \cos\left\{t(Y_i - b_0 - b_1 W_j) \right\}\right] dt \\
& - \int_0^{t^*} 2t^2 K_{t^*}(t) 
\left[\sum_i \sum_j \sin\left\{t(Y_i - b_0 - b_1 W_j) \right\}\right]\left[\sum_i \sum_j W_j \sin\left\{t(Y_i - b_0 - b_1 W_j) \right\}\right] dt , 
\end{split}
\]

and

\[
\begin{split}
\dfrac{\partial^2 D}{\partial b_1^2} = & 
\int_0^{t^*} 2t^2 K_{t^*}(t) 
\left[\sum_i \sum_j W_j \cos\left\{t(Y_i - b_0 - b_1 W_j) \right\}\right]^2 dt \\
& - \int_0^{t^*} 2t^2 K_{t^*}(t) 
\left[\sum_i \sum_j \sin\left\{t(Y_i - b_0 - b_1 W_j) \right\}\right]\left[\sum_i \sum_j W_j^2 \sin\left\{t(Y_i - b_0 - b_1 W_j) \right\}\right] dt.
\end{split}
\]
These expressions can be used to calculate the matrix $\hat{B}$ required for the bootstrap plug-in method for standard error estimation.

\section{A brief review of the Generalized Method of Moments}

In this section, we provide a brief overview of the generalized method of moments (GMM) approach to linear errors-in-variables models. GMM is a popular approach to estimating the parameters of linear EIV models. Recall the model 
\begin{equation*}
Y_i = \beta_0 + \beta_1 X_i + \varepsilon_i \quad\text{and}\quad W_i = X_i +U_i,
\end{equation*} 
$i=1,\ldots,n$. In this model, the parameters $\beta_0$ and $\beta_1$ are identifiable using moments of $W$ and $Y$ up to order $3$, provided $E[(X-\mu_X)^3]\neq 0$. Similarly, the parameters are identifiable using moments up to order $4$ provided the distributions of $X$, $U$, and $\varepsilon$ are not all Gaussian. We briefly review implementation of GMM here. Our approach is similar to that proposed by \cite{erickson2002two}. GMM is a viable nonparametric alternative to the phase function approach, in that no parametric model assumptions are required for implementation.

For GMM using sample moments up to order $K$, it is assumed that each of the variables $X$, $U$, and $\varepsilon$ has at least $2K$ finite moments. Furthermore, it is assumed that $U$ and $\varepsilon$ have distributions symmetric about $0$, $\mathrm{E}\left[U^{2k-1}\right] = \mathrm{E}\left[\varepsilon^{2k-1}\right] = 0$ for $k=1,2,\ldots,K$. Note that the use of the first $K$ moments requires that the underlying distributions have $2K$ moments for the estimators derived here to be asymptotically normally distributed with finite variance. 

Let $\mu_X$ denote the mean of $X$, and let $\sigma_X^2$, $\sigma_U^2$, and $\sigma_{\varepsilon}^2$ denote the respective variances of $X$, $U$, and $\varepsilon$. Additionally, let $\mu_{X,j} = \mathrm{E}\left[(X-\mu_X)^j\right]$ denote the $j$th centered moment of $X$, $j=3,\ldots,2K$, with equivalent notation holding for $\mu_{U,j}$ and $\mu_{\varepsilon,j}$. Finally, for the pair of random variables $(W,Y)$, let $\nu_{j,k}$ denote the joint centered moments,
\begin{equation}
\nu_{j,k} = \mathrm{E}\left[\left( W-\mu_{X}\right) ^{j}\left( Y-\beta _{0}-\beta _{1}\mu _{X}\right) ^{k}\right].
\label{joint_moments}
\end{equation}
Due to the independence of $X$, $U$, and $\varepsilon$, the joint moment $\nu_{j,k}$ can be expressed in terms of the marginal moments of $X$, $U$, and $\varepsilon$ up to order $j+k$. Making a few special cases explicit, note that $\nu_{2,0} = \sigma_X^2 + \sigma_U^2$, $\nu_{1,1} = \beta_1\sigma_X^2$, and $\nu_{0,2} = \beta_1^2\sigma_X^2 + \sigma_\varepsilon^2$. 

Now, let $\bm{\theta}_{(1)}=\left\{\mu_{X},\beta_{0},\beta_{1}\right\} $ and $\bm{\theta}_{(2)}=\left\{\sigma_X^2,\sigma_U^2,\sigma_\varepsilon^2\right\} $, and let $\bm{\theta}_{(2j-1)}=\left\{\mu_{X,2j-1}\right\} $ and $\bm{\theta}_{(2j)}=\left\{\mu_{X,2j},\mu_{U,2j},\mu_{\varepsilon,2j}\right\} $, $j=2,\ldots,\lfloor K/2\rfloor$, denote the higher-order moments.
Finally, let $\bm{\theta}_K=\left\{\bm{\theta}_{(1)},\ldots,\bm{\theta}_{(K)}\right\}$ denote the collection of unknown parameters required to specify a model up to order $K$.  The random variables
\[
A_{jk}\left( \bm{\theta }\right) =n^{-1/2}\sum_{i=1}^{n} \left\{ \left( W_{i}-\mu
_{X}\right) ^{j}\left( Y_{i}-\beta _{0}-\beta _{1}\mu _{X}\right) ^{k}-\nu
_{jk}\right\} 
\]%
have $\mathrm{E}\left[ A_{jk}\right] =0$ and $\mathrm{Cov}\left[ A_{jk},A_{j^{\prime },k^{\prime }}\right] = \nu _{j+j^{\prime },k+k^{\prime }}-\nu _{jk}\nu _{j^{\prime },k^{\prime }}$ for $j+j'+k+k'\leq 2K$. A such, the $A_{j,k}$ can be used to construct GMM estimators of the parameters. Specifically, let $\bm{A}_K(\bm{\theta}_K)$ denote the vector consisting of all terms $A_{jk}$ with $j,k=0,\ldots,K$ and $1\leq j+k \leq K$. Now, define $\bm{\Sigma}_K$ to be the covariance matrix corresponding to vector $\bm{A}_K$. This covariance matrix can be estimated empirically by defining joint sample moments \[\hat{\nu}_{j,k} = \frac{1}{n}\sum_{i=1}^{n} \left(W_i - \bar{W}\right)^j \left(Y_i - \bar{Y}\right)^k \] and subsequently letting \[\widehat{\mathrm{Cov}} \left[ A_{jk},A_{j^{\prime },k^{\prime }}\right] = \hat{\nu} _{j+j^{\prime },k+k^{\prime }}-\hat{\nu} _{jk}\hat{\nu} _{j^{\prime },k^{\prime }},\quad j+j'+k+k' \leq 2K. \] Let $\hat{\bm{\Sigma}}_K$ denote this estimated covariance matrix. The GMM parameter estimates are then found by minimizing the quadratic form
\begin{equation}
G_K(\bm{\theta}_K) = \bm{A}_K(\bm{\theta}_K)^\top \hat{\bm{\Sigma}}_K^{-1} \bm{A}_K(\bm{\theta}_K).
\label{GMM_eq}
\end{equation}

Note that the implementation of the GMM approach requires the use of $K \geq 3$, as the choices $K=1,2$ result in an overidentified system in terms of the parameters in $\bm{\theta}_K$.

\section{Additional Simulation Results}
\subsection{The effect of weighting function}
The simulation study in Section 5.1 (univariate EIV model) of the main paper explore three different choice of weighting function in calculating the phase function estimator: $K_1(t)=(1-|t|)^2\mathbb{I}(|t|\leq 1), ~ K_2(t)=(1-\vert t\vert)\mathbb{I}(|t|\leq 1) , ~ K_3(t)=(1-t^2)\mathbb{I}(|t|\leq 1)$. Table C.1 presents the median SE of phase function estimates for these three weight function choices for a subset of simulation settings with $X \sim |N(0,1)|$ or $X\sim \exp(1)$, and $(p_W, p_Y) = (0.25,0.40)$. The results for simulation configurations not reported in Table C.1 follow the same general patterns.

\begin{table}[H]
	\centering
	\caption{Median square error and the corresponding interquartile range, scaled by the sample size, for the phase function estimators with weighting functions $K_1(t), ~ K_2(t)$, and $K_3(t)$.}
	\begin{tabular}{rrr|cccccc}
		\hline
		True X & $n$ & Error &  \multicolumn{2}{c}{$K_1(t)$} & \multicolumn{2}{c}{$K_2(t)$} & \multicolumn{2}{c}{$K_3(t)$} \\
		\cline{4-9} 
		&& & $\beta_0$ & $\beta_1$ & $\beta_0$ & $\beta_1$& $\beta_0$ & $\beta_1$\\ 
		\hline
		$\vert N(0,1) \vert$ & 500 & Normal & 3.37 & 4.41 & 3.47 & 4.79 & 3.4 & 4.74 \\ 
		&  &  & (0.14) & (0.18) & (0.14) & (0.19) & (0.14) & (0.19) \\ 
		&& Laplace & 1.9 & 3.19 & 1.83 & 3.07 & 1.84 & 3.04 \\ 
		&  &  & (0.09) & (0.14) & (0.08) & (0.13) & (0.08) & (0.13) \\ 
		& 1000 & Normal & 3.21 & 4.38 & 3.31 & 4.6 & 3.26 & 4.44 \\ 
		&  &  & (0.14) & (0.18) & (0.14) & (0.2) & (0.14) & (0.19) \\ 
		&& Laplace & 1.65 & 3.01 & 1.62 & 2.88 & 1.62 & 2.89 \\ 
		&  &  & (0.08) & (0.13) & (0.08) & (0.12) & (0.08) & (0.12) \\ 
		\hline
		$\exp(1)$ & 500 & Normal & 4.75 & 4.28 & 5.08 & 5.07 & 4.94 & 4.72 \\ 
		&  &  & (0.21) & (0.2) & (0.23) & (0.24) & (0.23) & (0.23) \\ 
		&& Laplace & 2.5 & 3.63 & 2.59 & 3.89 & 2.52 & 3.63 \\ 
		&  &  & (0.11) & (0.16) & (0.12) & (0.18) & (0.11) & (0.17) \\ 
		& 1000 & Normal & 5.55 & 4.97 & 5.92 & 6.04 & 5.86 & 5.49 \\ 
		&  &  & (0.24) & (0.23) & (0.26) & (0.27) & (0.25) & (0.25) \\ 
		&& Laplace & 2.62 & 3.2 & 2.56 & 3.68 & 2.56 & 3.29 \\ 
		&  &  & (0.11) & (0.15) & (0.12) & (0.17) & (0.11) & (0.15) \\ 
		\hline
	\end{tabular}	
	\label{tab:wt}		
\end{table}

As can be seen in Table \ref{tab:wt}, the choice of weights function does not have a major impact on the quality of the estimators when using median square error as criterion. However, the choice of weight function $K_1(t)=(1-|t|)^2\mathbb{I}(|t|\leq 1)$ most often results in the lowest median square error for both $\beta_0$ and $\beta_1$. As such, the phase function-based estimators are
compared to the other methods of estimation for this choice of weight function.
\subsection{Full Simulation Results for Univariate EIV model}
In this section, we present the full results for the simulation studies in the simple EIV setting in Section $5.1$ of the main paper. Data were generated according to the model
$Y_i = \beta_0 + \beta_1 X_i + \varepsilon_i$ and $W_i = X_i + U_i$, $i= 1,\ldots, n$ with true parameters $(\beta_0,\beta_1) = (1,3)$. Three asymmetric distributions were used to simulate $X$, namely (1) a half-normal distribution, $X \sim |N(0,1)|$, (2) an exponential distribution, $X \sim \exp(1)$, and (3) a bimodal mixture distribution, $X \sim 0.5N(5,1^2)+0.5N(2.5,0.6^2)$. Three different distributions were considered for error components $U$ and $\varepsilon$, namely the normal, $t$-distribution with $2.5$ degrees of freedom, and the Cauchy distribution. For the Normal and $t_{2.5}$ distributions, the error components were simulated to have mean $0$ and respective variances $\sigma_U^2$ and $\sigma_\varepsilon^2$. For the Cauchy distribution, the error components were simulated to be symmetric about $0$ and have respective interquartile range (IQR) $\sigma_U$ and $\sigma_\varepsilon$. The variance and IQR parameters were chosen to achieve specific noise-to-signal ratios, $p_W = \sigma_U^2 / \sigma_X^2$ and $p_Y = \sigma_{\varepsilon}^2 / (\beta_1 \sigma_X)^2$. The noise-to-signal ratios pairs reported here are $ (p_W,p_Y)\in \left\{(0.075,0.15),(0.25, 0.40)\right\}$. Results are reported for sample sizes $n \in \{500, 1000\}$. For each configuration, the median square error of each estimator is reported with the corresponding interquartile range. 
\begin{table}[H]
	\centering
	\caption{Median square errors of estimators and the corresponding interquartile range (in parentheses), scaled by the sample size, in the univariate regression simulation when the true distribution of $X$ is half-normal.}
	
	\begin{tabular}{rlc| cc cc cc HH cc}
		\hline
		$n$ & Error & ($p_W, p_Y$) &  \multicolumn{2}{c}{Naive} & \multicolumn{2}{c}{GMM} & \multicolumn{2}{p{2cm}}{Disattenuation} & & & \multicolumn{2}{c}{Phase function} \\
		\cline{4-13}	
		&&&  $\beta_0$ & $\beta_1$ & $\beta_0$ & $\beta_1$& $\beta_0$ & $\beta_1$ & $\beta_0$ & $\beta_1$ & $\beta_0$ & $\beta_1$\\ 
		\hline
		500 & Normal & (0.075,0.15) & 14.16 & 21.94 & 1.17 & 1.58 & 0.5 & 0.53 & 1.16 & 1.48 & 0.95 & 1.27 \\ 
		&  &  & (0.16) & (0.21) & (0.05) & (0.07) & (0.02) & (0.02) & (0.05) & (0.07) & (0.04) & (0.05) \\ 
		& & (0.25,0.40) & 114.89 & 180.01 & 3.69 & 5.12 & 1.57 & 1.82 & 3.75 & 5 & 3.37 & 4.41 \\ 
		&  &  & (0.73) & (0.97) & (0.18) & (0.24) & (0.07) & (0.09) & (0.18) & (0.24) & (0.14) & (0.18) \\ 
		& $t_{2.5}$ & (0.075,0.15) & 8.45 & 13.18 & 0.78 & 1.11 & 1.45 & 2.16 & 1.21 & 1.59 & 0.59 & 1.01 \\ 
		&  &  & (0.16) & (0.22) & (0.03) & (0.05) & (0.06) & (0.07) & (0.06) & (0.08) & (0.03) & (0.04) \\ 
		& & (0.25,0.40) & 71.86 & 112.03 & 3.67 & 5.34 & 12.2 & 18.13 & 6.23 & 8.63 & 1.9 & 3.19 \\ 
		&  &  & (0.95) & (1.39) & (0.2) & (0.28) & (0.38) & (0.55) & (0.33) & (0.51) & (0.09) & (0.14) \\ 
		1000 & Normal & (0.075,0.15) & 28.39 & 44.16 & 1.17 & 1.62 & 0.53 & 0.56 & 1.2 & 1.61 & 0.94 & 1.21 \\ 
		&  &  & (0.24) & (0.3) & (0.06) & (0.08) & (0.02) & (0.02) & (0.06) & (0.07) & (0.04) & (0.06) \\ 
		& & (0.25,0.40) & 229.6 & 358.97 & 4.45 & 5.97 & 1.75 & 1.89 & 4.36 & 5.87 & 3.21 & 4.38 \\ 
		&  &  & (1.1) & (1.39) & (0.19) & (0.24) & (0.08) & (0.08) & (0.18) & (0.24) & (0.14) & (0.18) \\ 
		& $t_{2.5}$ & (0.075,0.15) & 18.54 & 29.12 & 0.98 & 1.31 & 2.05 & 3.08 & 1.51 & 2.26 & 0.63 & 0.99 \\ 
		&  &  & (0.27) & (0.39) & (0.05) & (0.06) & (0.08) & (0.1) & (0.08) & (0.11) & (0.03) & (0.04) \\ 
		& & (0.25,0.40) & 154.67 & 243.3 & 5.04 & 7.44 & 17.67 & 26.1 & 9.49 & 13.95 & 1.65 & 3.01 \\ 
		&  &  & (1.67) & (2.52) & (0.3) & (0.47) & (0.52) & (0.79) & (0.56) & (0.81) & (0.08) & (0.13) \\ 
		\hline
	\end{tabular}
	\label{tab:mse_hn}
\end{table}

\begin{table}[H]
	\centering
	\caption{Median square errors of estimators and the corresponding interquartile range (in parentheses), scaled by the sample size, in the univariate regression simulation when the true distribution of $X$ is exponential.}
	
	\begin{tabular}{rlc| cc cc cc HH cc}
		\hline
		n & Error & NSR &  \multicolumn{2}{c}{Naive} & \multicolumn{2}{c}{GMM} & \multicolumn{2}{p{2cm}}{Disattenuation} & & & \multicolumn{2}{c}{Phase function} \\
		\cline{4-13}	
		&&&  $\beta_0$ & $\beta_1$ & $\beta_0$ & $\beta_1$& $\beta_0$ & $\beta_1$ & $\beta_0$ & $\beta_1$ & $\beta_0$ & $\beta_1$\\
		\hline
		500 & Normal & (0.075,0.15) & 22.25 & 22.4 & 1.46 & 0.95 & 1.2 & 0.67 & 1.39 & 0.96 & 1.62 & 1.69 \\ 
		&  &  & (0.31) & (0.24) & (0.07) & (0.05) & (0.05) & (0.03) & (0.07) & (0.05) & (0.07) & (0.07) \\ 
		&& (0.25,0.40) & 178.85 & 181.39 & 4.77 & 3.08 & 4.04 & 2.69 & 4.67 & 2.95 & 4.75 & 4.28 \\ 
		&  &  & (1.33) & (1.17) & (0.21) & (0.15) & (0.17) & (0.11) & (0.21) & (0.15) & (0.21) & (0.2) \\ 
		& $t_{2.5}$ & (0.075,0.15) & 13.19 & 13.39 & 0.95 & 0.58 & 2.57 & 2.3 & 1.57 & 1.15 & 0.88 & 1.07 \\ 
		&  &  & (0.29) & (0.24) & (0.04) & (0.03) & (0.1) & (0.08) & (0.08) & (0.06) & (0.04) & (0.05) \\ 
		& & (0.25,0.40) & 117.47 & 113.1 & 2.97 & 2.26 & 17.69 & 19.25 & 6.62 & 5.62 & 2.5 & 3.63 \\ 
		&  &  & (1.69) & (1.59) & (0.14) & (0.11) & (0.65) & (0.58) & (0.37) & (0.31) & (0.11) & (0.16) \\ 
		\hline
		1000 & Normal & (0.075,0.15) & 43.83 & 44.13 & 1.55 & 1.07 & 1.16 & 0.67 & 1.46 & 1.08 & 1.74 & 1.72 \\ 
		&  &  & (0.44) & (0.34) & (0.07) & (0.05) & (0.05) & (0.03) & (0.07) & (0.05) & (0.08) & (0.07) \\ 
		&& (0.25,0.40) & 363.62 & 362.91 & 5.18 & 3.48 & 4.08 & 2.8 & 5.21 & 3.53 & 5.55 & 4.97 \\ 
		&  &  & (1.99) & (1.68) & (0.23) & (0.15) & (0.19) & (0.12) & (0.23) & (0.15) & (0.24) & (0.23) \\ 
		& $t_{2.5}$ & (0.075,0.15) & 29.74 & 28.94 & 1.05 & 0.65 & 3.28 & 2.96 & 2.02 & 1.51 & 0.92 & 1.27 \\ 
		&  &  & (0.47) & (0.41) & (0.05) & (0.04) & (0.13) & (0.11) & (0.1) & (0.08) & (0.04) & (0.06) \\ 
		&& (0.25,0.40) & 249.07 & 242.99 & 4.28 & 3.06 & 27.39 & 27.79 & 8.68 & 6.56 & 2.62 & 3.2 \\ 
		&  &  & (2.89) & (2.67) & (0.19) & (0.15) & (0.93) & (0.86) & (0.44) & (0.4) & (0.11) & (0.15) \\ 
		\hline
	\end{tabular}
	\label{tab:mse_exp}
\end{table}

\begin{table}[H]
	\centering
	\caption{Median square errors of estimators and the corresponding interquartile range (in parentheses), scaled by the sample size,in the univariate regression simulation when the true distribution of $X$ is a mixture of normal distributions.}
	
	\begin{tabular}{rlc| cc cc cc HH cc}
		\hline
		n & Error & NSR &  \multicolumn{2}{c}{Naive} & \multicolumn{2}{c}{GMM} & \multicolumn{2}{p{2cm}}{Disattenuation} &&& \multicolumn{2}{c}{Phase function} \\
		\cline{4-13}	
		&&&  $\beta_0$ & $\beta_1$ & $\beta_0$ & $\beta_1$& $\beta_0$ & $\beta_1$ & $\beta_0$ & $\beta_1$ & $\beta_0$ & $\beta_1$\\ 
		\hline
		500 & Normal & (0.075,0.15) & 306.68 & 21.62 & 63.48 & 4.74 & 9.16 & 0.53 & 62.93 & 4.65 & 23.56 & 1.61 \\ 
		&  &  & (3.31) & (0.21) & (2.52) & (0.17) & (0.39) & (0.02) & (2.49) & (0.17) & (1.14) & (0.08) \\ 
		&& (0.25,0.40) & 2528.79 & 178.52 & 254.05 & 17.4 & 25.12 & 1.51 & 247.56 & 17.06 & 143.37 & 9.76 \\ 
		&  &  & (13.87) & (0.87) & (11.5) & (0.8) & (1.14) & (0.07) & (11.04) & (0.77) & (6.73) & (0.48) \\ 
		& $t_{2.5}$ & (0.075,0.15) & 192.04 & 13.71 & 43.9 & 2.97 & 28.71 & 2.09 & 45.56 & 3.08 & 15.48 & 0.96 \\ 
		&  &  & (3.21) & (0.22) & (1.45) & (0.1) & (1.05) & (0.08) & (2.04) & (0.14) & (0.71) & (0.05) \\ 
		&& (0.25,0.40) & 1572.13 & 111.86 & 314.14 & 22.16 & 249.82 & 18.35 & 290.67 & 19.8 & 67.16 & 4.39 \\ 
		&  &  & (20.39) & (1.37) & (10.69) & (0.74) & (7.76) & (0.55) & (14.47) & (0.97) & (3.43) & (0.23) \\ 
		\hline
		1000 & Normal & (0.075,0.15) & 616.32 & 43.87 & 71.47 & 4.99 & 9.1 & 0.55 & 70.38 & 4.98 & 20.96 & 1.41 \\ 
		&  &  & (4.67) & (0.3) & (3.29) & (0.23) & (0.38) & (0.02) & (3.3) & (0.23) & (0.97) & (0.06) \\ 
		&& (0.25,0.40) & 5047.12 & 361.45 & 245.51 & 17.2 & 25.21 & 1.54 & 249.66 & 17.34 & 101.42 & 7.29 \\ 
		&  &  & (19.57) & (1.24) & (10.93) & (0.75) & (1.1) & (0.07) & (10.93) & (0.75) & (4.91) & (0.35) \\ 
		& $t_{2.5}$ & (0.075,0.15) & 403.52 & 28.6 & 77.56 & 5.38 & 44.06 & 3.11 & 78.17 & 5.42 & 14.57 & 0.94 \\ 
		&  &  & (5.49) & (0.39) & (2.61) & (0.18) & (1.57) & (0.11) & (3.34) & (0.23) & (0.64) & (0.04) \\ 
		& & (0.25,0.40) & 3474.56 & 246.26 & 546.55 & 38.33 & 373.97 & 27.02 & 532.79 & 35.78 & 45.81 & 3.05 \\ 
		&  &  & (35.62) & (2.5) & (19.73) & (1.37) & (11.8) & (0.82) & (23.81) & (1.65) & (2.52) & (0.16) \\ 
	\end{tabular}
	\label{tab:mse_bimodal}
\end{table}	
\begin{table}[H]
	\centering
	\caption{Median square error and interquartile range of the GMM and phase function estimators in the univariate regression simulation when model errors are Cauchy}
	\begin{tabular}{lrc|cccc}
		\hline
		True X& $n$ & $(p_W, p_Y)$ & \multicolumn{2}{c}{GMM} & \multicolumn{2}{c}{Phase function} \\ 
		\cline{4-7}
		&& & $\beta_0$ & $\beta_1$ & $\beta_0$ & $\beta_1$ \\
		\hline
		$\vert N(0,1) \vert$ & 500 & (0.075,0.15) & 4.04 & 8.97 & 0.02 & 0.04 \\ 
		&  &  & (0.07) & (0.07) & (0.00) & (0.00) \\ 
		&& (0.25,0.40) & 4.44 & 8.99 & 0.05 & 0.1 \\ 
		&  &  & (0.07) & (0.02) & (0.00) & (0.00) \\ 
		& 1000 & (0.075,0.15) & 4.5 & 9 & 0.01 & 0.02 \\ 
		&  &  & (0.07) & (0.03) & (0.00) & (0.00) \\ 
		&& (0.25,0.40) & 4.42 & 9 & 0.02 & 0.04 \\ 
		&  &  & (0.07) & (0.01) & (0.00) & (0.00) \\ 
		\hline
		$\exp(1)$ & 500 & (0.075,0.15) & 4.09 & 8.92 & 0.01 & 0.02 \\ 
		&  &  & (0.11) & (0.12) & (0.00) & (0.00) \\ 
		&& (0.25,0.40) & 5.43 & 8.99 & 0.02 & 0.05 \\ 
		&  &  & (0.12) & (0.05) & (0.00) & (0.00) \\ 
		& 1000 & (0.075,0.15) & 5.2 & 8.98 & 0 & 0.01 \\ 
		&  &  & (0.12) & (0.08) & (0.00) & (0.00) \\ 
		& & (0.25,0.40) & 6.29 & 9 & 0.01 & 0.02 \\ 
		&  &  & (0.12) & (0.02) & (0.00) & (0.00) \\ 
		\hline
		Bimodal & 500 & (0.075,0.15) & 19.71 & 8.75 & 2.12 & 0.13 \\ 
		&  &  & (1.8) & (0.13) & (0.09) & (0.01) \\ 
		& & (0.25,0.40) & 52.39 & 8.94 & 2.32 & 0.15 \\ 
		&  &  & (1.89) & (0.07) & (0.11) & (0.01) \\ 
		&1000 & (0.075,0.15) & 29.02 & 8.93 & 1.34 & 0.08 \\ 
		&  &  & (1.89) & (0.08) & (0.06) & (0.00) \\ 
		& & (0.25,0.40) & 53.6 & 8.98 & 1.85 & 0.12 \\ 
		&  &  & (1.89) & (0.02) & (0.08) & (0.01) \\ 
		\hline
	\end{tabular}
	
\end{table}	

\subsection{Full Simulation Results for Multiple Regression}
In this section, we present the full results for the simulation study in the multiple EIV linear model setting in the section 5 of the main paper. Data were simulated according to the model $Y_i = \beta_0 + \beta_X X_i + \beta_Z Z_i + \epsilon_i$, $W_i = X_i + U_i$, $i=1,\ldots, n$ with parameters $\beta_0 = 0, \beta_X = 3$, and $\beta_Z = 2$. Here, $X$ is the error-prone covariate while $Z$ is error-free. Samples sizes $n \in \{1000, 2000\}$ were considered. We include here results for the two cases $X$ half-normal and $X$ having the bimodal normal mixture defined in the simple EIV setting. The covariate $Z$ was generated from the same distribution as $X$, and a normal copula with $\rho = 0.5$ was used to generate $X$ and $Z$ to be correlated. The error distributions were taken to be normal and Laplace with noise-to-signal ratios as in the simple EIV model. For each simulation configuration, 2000 replications were run. Table \ref{tab:mse_multtiple_2} and \ref{tab:mse_multtiple_bimodal} presents the median square error for the naive, phase function, and SIMEX estimator with their corresponding interquartile ranges.  

\begin{table}[H]
	\centering
	\caption{Median square error and interquartile range (in parentheses), scaled by the sample size for the estimators in the multivariate regression simulation when $X$ and $Z$ are half-normal and correlated with correlation $\rho=.5$.}
	
	\begin{tabular}{ccc|cccccc}
		\hline
		$n$  & Error          & $(p_W, p_Y)$    & \multicolumn{2}{c}{Naive} & \multicolumn{2}{c}{Phase function}  & \multicolumn{2}{c}{SIMEX} 
		\\ 
		\cline{4-9}
		&&& $\beta_X$ & $\beta_Z$ & $\beta_X$ & $\beta_Z$& $\beta_X$ & $\beta_Z$ \\
		\hline
		1000 & Normal & (0.075,0.15) & 715.69 & 167.7 & 7.34 & 9.96 & 4.19 & 2.78 \\ 
		&  &  & (1.92) & (1.01) & (0.32) & (0.46) & (0.18) & (0.13) \\ 
		& & (0.25,0.40) & 2349.55 & 540.65 & 328.67 & 613.13 & 21.94 & 15.26 \\ 
		&  &  & (4.31) & (2.57) & (148.75) & (71.23) & (0.97) & (0.66) \\ 
		& Laplace & (0.075,0.15) & 705.18 & 164.94 & 7.75 & 12.76 & 5.79 & 3.75 \\ 
		&  &  & (2.15) & (1.05) & (0.38) & (0.58) & (0.27) & (0.16) \\ 
		& & (0.25,0.40) & 2329.48 & 539.89 & 331.13 & 658.7 & 32.09 & 16.73 \\ 
		&  &  & (4.73) & (2.55) & (148.71) & (71.53) & (1.48) & (0.74) \\ 
		\hline
		2000 & Normal & (0.075,0.15) & 1417.18 & 326.47 & 7.13 & 9.26 & 4.19 & 3.1 \\ 
		&  &  & (2.73) & (1.44) & (0.3) & (0.43) & (0.19) & (0.15) \\ 
		& & (0.25,0.40) & 4691.1 & 1089.42 & 46.83 & 86.77 & 21.41 & 13.41 \\ 
		&  &  & (5.95) & (3.82) & (2.86) & (5.69) & (0.96) & (0.61) \\ 
		& Laplace & (0.075,0.15) & 1419.04 & 326.72 & 8.34 & 12.18 & 6.12 & 3.32 \\ 
		&  &  & (3.48) & (1.55) & (0.36) & (0.57) & (0.27) & (0.16) \\ 
		& & (0.25,0.40) & 4679.33 & 1073.46 & 51.75 & 109.01 & 36.64 & 17.17 \\ 
		&  &  & (7.36) & (3.86) & (3.22) & (7.79) & (1.51) & (0.8) \\ 
		\hline
	\end{tabular}
	\label{tab:mse_multtiple_2}
\end{table}
\begin{table}[h!]
	\centering
	\caption{Median square error and interquartile range (in parentheses), scaled by the sample size, for the estimators in the multivariate regression simulation when $X$ and $Z$ are mixtures of normal distribution and correlated with correlation $\rho=.5$.}
	
	\begin{tabular}{ccc|cccccc}
		\hline
		$n$  & Error          & $(p_W, p_Y)$    & \multicolumn{2}{c}{Naive} & \multicolumn{2}{c}{Phase function}  & \multicolumn{2}{c}{SIMEX} 
		\\ 
		\cline{4-9}
		&&& $\beta_X$ & $\beta_Z$ & $\beta_X$ & $\beta_Z$& $\beta_X$ & $\beta_Z$ \\
		\hline
		1000 & Normal & (0.075,0.15) & 241.69 & 56.84 & 13.44 & 18.65 & 1.55 & 1.29 \\ 
		&  &  & (0.83) & (0.45) & (0.56) & (0.85) & (0.07) & (0.06) \\ 
		& & (0.25,0.40) & 1056.68 & 249.27 & 69.64 & 117.72 & 6.91 & 4.81 \\ 
		&  &  & (2.22) & (1.47) & (5.55) & (9.45) & (0.29) & (0.23) \\ 
		& Laplace & (0.075,0.15) & 239.77 & 56.15 & 14.74 & 24.13 & 2.04 & 1.62 \\ 
		&  &  & (0.96) & (0.47) & (0.74) & (1.23) & (0.09) & (0.07) \\ 
		&& (0.25,0.40) & 1056.6 & 248.34 & 149.4 & 250.81 & 9.49 & 5.65 \\ 
		&  &  & (2.86) & (1.45) & (19.69) & (65.43) & (0.46) & (0.26) \\ 
		\hline 
		2000 & Normal & (0.075,0.15) & 483.2 & 113.82 & 10.74 & 14.71 & 1.77 & 1.5 \\ 
		&  &  & (1.13) & (0.63) & (0.46) & (0.71) & (0.07) & (0.06) \\ 
		&& (0.25,0.40) & 2127.88 & 498.96 & 55.95 & 94.55 & 6.2 & 4.77 \\ 
		&  &  & (3.16) & (2.02) & (2.87) & (4.63) & (0.26) & (0.22) \\ 
		& Laplace & (0.075,0.15) & 486.34 & 113.48 & 11.72 & 21.03 & 2.13 & 1.59 \\ 
		&  &  & (1.4) & (0.66) & (0.62) & (0.93) & (0.1) & (0.07) \\ 
		& & (0.25,0.40) & 2125.8 & 499.99 & 80.09 & 144.17 & 9.69 & 5.44 \\ 
		&  &  & (4.18) & (2.11) & (4.47) & (7.74) & (0.45) & (0.25) \\ 
		\hline
	\end{tabular}
	\label{tab:mse_multtiple_bimodal}
\end{table}

\section{Additional Data Examples}
\subsection{Abrasiveness Index Data}
The data analyzed here was originally considered by \cite{lombard2005nonparametric} in the context of estimating a quantile comparison function from paired data. Observations are pairs $(W_j,Y_j)$, $j=1,\ldots,98$, where both $W_j$ and $Y_j$ represent measures of the abrasiveness index (AI) of a batch of coal. The AI is considered a proxy for the quality of coal, and is used to determine the price of a batch of coal. The $Y_j$ measurements were obtained using the YGP method, see
\cite{yancey1951investigation}. This method is widely used, but is costly to implement. The $W_j$ measurements were obtained using a similar method that is less involved and cheaper to implement. Contracts are typically written in terms of the YGP measurements, and it is of interest to determine the relationship between the new method and the YGP method. Here, we treat both the $W_j$ and $Y_j$ data as contaminated versions of the true quality of a batch of coal, denoted $X_j$. Assume that the linear errors-in-variables structure holds, i.e. $W_j = X_j + U_j$ and $Y_j = \beta_0 + \beta_1X_j + \varepsilon_j$.

\begin{figure}
	\includegraphics[width=15cm]{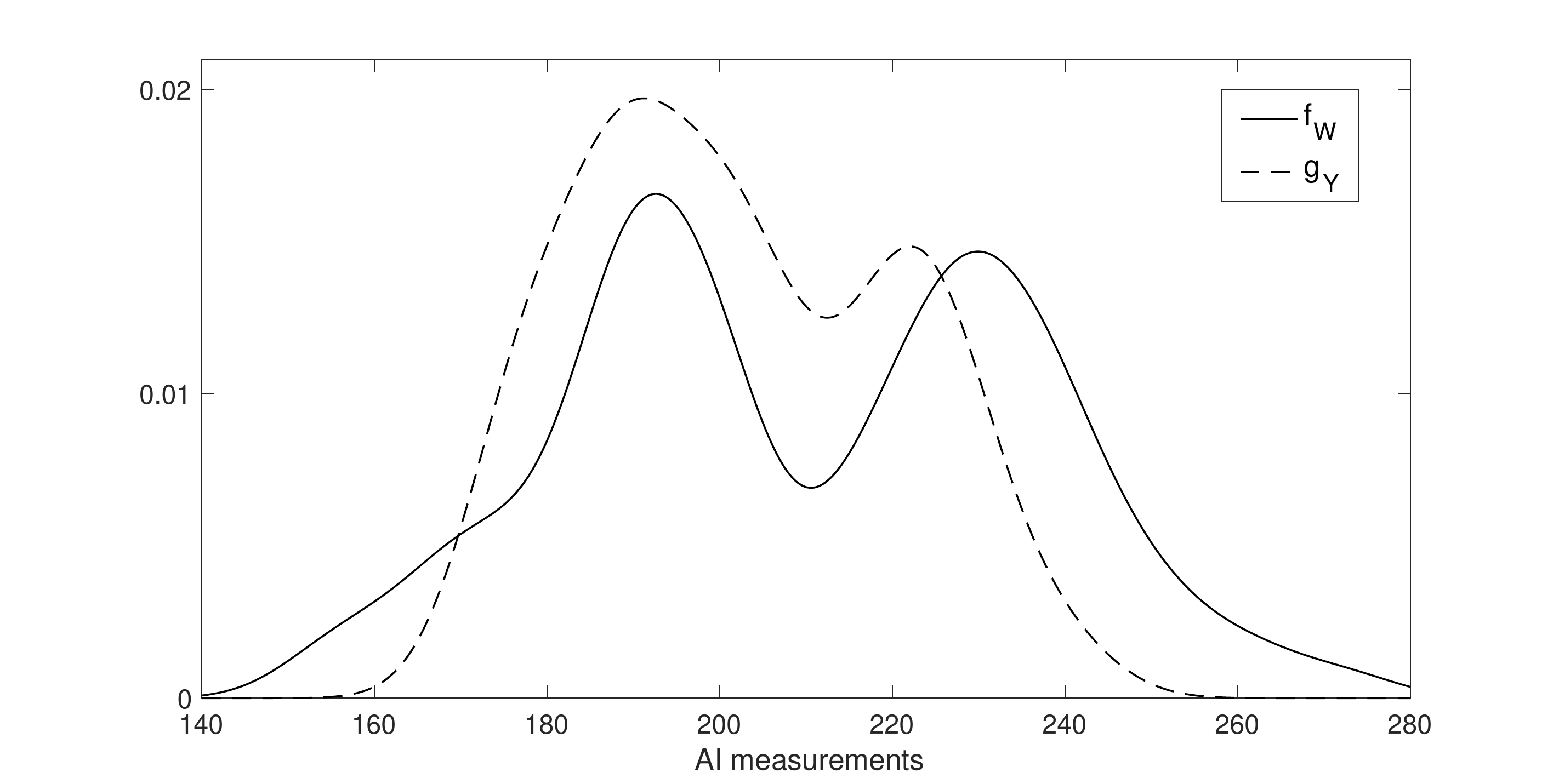}
	\caption{Kernel Density Estimators for $W$ (new method) and $Y$ (YGP method) data}
	\label{fig:AI_kerns}
\end{figure}

In Figure \ref{fig:AI_kerns}, we show kernel density estimates using normal reference plug-in bandwidths for both $W$ and $Y$. 
Bandwidths selected using unbiased cross-validation were also considered, but did not alter the estimates in a visually discernible way. For the given data, the naive regression estimators, GMM estimators, and phase function-based estimators were calculated; the results are reported in Table \ref{tab:AI_estims}. Also reported are the estimated variance components based on the second sample moments. Specifically, $\hat{\sigma}_X^2 = s_{WY}/\hat{\beta}_1$, $\hat{\sigma}_U^2 = \max\{0,s_W^2 - \hat{\sigma}_X^2\}$, and $\hat{\sigma}_\varepsilon^2 = \max\{0,s_Y^2 - \hat{\beta}_1^2\hat{\sigma}_X^2\}$, where $s_{WY}$ denotes the sample covariance, and $s_W^2$ and $s_Y^2$ denote the sample variances.

\begin{table}[h]
	\begin{center}
		\caption{Naive, GMM, and phase function-based estimators of the linear errors-in-variables for the abrasiveness index data.}
		\begin{tabular}{c|ccccc}
			\hline
			Method & $\hat{\beta}_0$ & $\hat{\beta}_1$ & $\hat{\sigma}_X^2$ & $\hat{\sigma}_U^2$ & $\hat{\sigma}_{\varepsilon}^2$ \\ \hline
			Naive  & 94.959          & 0.511           & 709.478            & 0                  & 242.123                        \\ 
			GMM    & 14.619          & 0.895           & 398.862            & 279.289            & 0                              \\ 
			Phase  & -40.776         & 1.157           & 313.066            & 396.411            & 0                              \\ \hline
		\end{tabular}
		\label{tab:AI_estims}
	\end{center}
\end{table}

The results in Table \ref{tab:AI_estims} are striking. As one would expect, the naive estimator of slope is shrunk towards $0$ when compared to the GMM and phase function estimators of slope. Both GMM and the phase function approach suggest that, as seen by the estimates of $\sigma_U^2$, the new method introduces a large amount of measurement error. On the other hand, the established YGP method has estimated measurement error $0$. Due to the small sample size, we are hesitant to conclude that the YGP method is error free. However, the results do suggest that if the YGP method does introduce measurement error, it is small relative to the measurement error introduced by the new method. Any company considering adoption of the new method for measuring the abrasiveness index should whether the increased measurement error is worth the cost savings of the new method.

To assess the variability of the computed estimators, pairwise bootstrap resampling was used. A total of $B=2000$ bootstrap samples were taken. Both GMM and the phase function method is prone to outliers in small samples. Subsequently, the interquartile ranges (IQR) of the respective bootstrap distributions were used as robust measures of spread. For GMM, $\mathrm{IQR}^*(\beta_0) = 71.167$ and $\mathrm{IQR}^*(\beta_1) = 0.335$. For the phase function method, $\mathrm{IQR}^*(\beta_0) = 56.031$ and $\mathrm{IQR}^*(\beta_1) = 0.271$. While this suggests that the phase function method gives less variable results, we should note that it is possible to choose a different measure of spread that contradicts this conclusion. Specifically, the difference between the $10$th and $90$th percentiles of the bootstrap distributions gives estimated spread $0.450$ and $0.624$ for the slope estimators using GMM and the phase function method respectively. Ultimately, for the data at hand, it is not possible to conclude that one method is superior to the other.

\subsection{Analysis of OPEN study}

In this section, the relationships between true dietary intakes and various measurements like biomarkers, diary, and self-report instruments are studied. In the National Cancer's Institute OPEN study, two indicators of dietary intakes of interest include protein intake and energy intake. For each indicator, each intake was measured by a food frequency questionnaire (FFQ), a 24-hour recall interview, and a biomarker. Each measurement is replicated twice. The dataset is used to illustrate several examples of measurement error modeling in \cite{carroll2006measurement}. The data made available on the website of the cited monograph is not the actual data from the OPEN Study, but has been simulated to have similar properties to the true data. These are $n=223$ observations in this dataset.    

For each indicator, the fitted model is of the form $Y_i = \beta_0+\beta_1X_{1i}+ \beta_2X_{2i} + \beta_3 X_{3i}+\varepsilon_i$, and $W_{jik} = X_{jik} + U_{jik},$ $j=1,2,3,$ $i=1,\ldots,n$, $k=1,2$ ,  where $Y_i$ is the true amount of the indicator, $X_{1i}, X_{2i}$ and $X_{3i}$ represent the (unobserved) amount of the indicator from biomarker measurement, FFQ, and interview of the $i^{th}$ subject respectively. If there is no measurement error exists, all the values $X_{1i}, X_{2i}$ and $X_{3i}$ would be equal to the value of $Y_i$. However, the observed data $W_{jik}$ are all different from $Y_i$, showing measurement error exists in all of the measurements.  

The estimators that are computed include the naive estimator, the simulation-extrapolation (SIMEX) estimator, and the phase function estimator. All the estimators are computed based on $Y_i$ and $W_{ji} = \dfrac{1}{2} (W_{ji1}+W_{ji2})$. Note that the SIMEX estimator requires knowledge of the variance of the measurement errors, which is possible to estimate in this situation because replication data for each measurement is available. The variance of measurement error associated with $W_{ji}$ was computed as
\[ \hat{\sigma}^2_j = \dfrac{1}{2n(n-1)} \sum_{i=1}^{n} (W_{ji1}-W_{ji2})^2.
\]
The phase function estimator was computed by minimizing the statistic 
\[D = \int_{0}^{t^*} \left(\sum_{i=1}^{n} \sum_{j=1}^{n} \sin\left[t\left(Y_j- W_{1i} \beta_{1i}-W_{2i} \beta_{2i}-W_{3i} \beta_{3i}\right)\right]\right)^2K_{t^*}(t) dt.
\]
with $K(t)=(1-|t|)^2$ and $t^*$ being the smallest $t>0$ such that $\vert \hat{\phi}_Y(t)\vert \leq n^{-1/4}$. This minimization problem is nonconvex, so the numerical algorithm was started at numerous points around the naive estimate. The estimates and its estimated standard error  (in parentheses) for both protein and energy intake were given in the Table \ref{tab:protein} and \ref{tab:energy}  below. The standard error for the phase function and the SIMEX estimates was  computed to be the interquartile range (IQR) of the corresponding estimates from $B=100$ bootstrap samples, while the standard error for the naive was computed using the traditional Fisher information matrix.

\begin{table}[h]
	\centering
	\caption{Analysis of simulated OPEN data for protein intake} 
	
	\begin{tabular}{cccc}
		Measurement &Naive & SIMEX & Phase Function \\
		\hline
		FFQ &0.041 (0.022)&0.194 (0.068) &-0.072 (0.306)\\
		24-hour recall &0.041 (0.022)&0.051 (0.036)& 0.127 (0.243)\\
		Biomarker &0.587 (0.037)&1.018 (0.138) &0.836 (0.400) 
	\end{tabular}	
	\label{tab:protein}
\end{table}

\begin{table}[h]
	\centering
	\caption{Analysis of simulated OPEN data for energy intake} 
	
	\begin{tabular}{cccc}
		Measurement &Naive & SIMEX & Phase Function \\
		\hline
		FFQ &0.006 (0.008)& 0.003 (0.022)&0.133 (0.148)\\
		24-hour recall & 0.006 (0.010)&0.007 (0.037)& 0.226 (0.256) \\
		Biomarker & 0.932 (0.017)& 0.986 (0.028) & 0.859 (0.268)
	\end{tabular}	
	\label{tab:energy}
\end{table}

The results from Table \ref{tab:protein} and \ref{tab:energy} show that for both protein and energy intake, only biomarker measurements have significant effect on the true amount. In the case of protein intake, the naive estimates attentuates the effect of the biomarker considerably, while the SIMEX and phase function estimates are able to correct it. In the case of energy intake, the phase function estimate reduces the magnitude of the relationship between biomarker measurement and the true amount. Compared to the SIMEX estimate, the phase function estimator has a much higher standard error. This is expected because the SIMEX estimator uses knowledge of the measurement error variances, while the phase function estimator does not. 

\end{document}